\documentclass[a4paper,11pt]{article}

\usepackage{lineno,hyperref}
\DeclareMathAlphabet{\pazocal}{OMS}{zplm}{m}{n}

\usepackage{dsfont}
\usepackage{amsmath}
\usepackage{amssymb}
\usepackage{tcolorbox}
\usepackage{subfig}
\usepackage{float}
\usepackage{longtable}
\usepackage{booktabs}
\usepackage{lscape}
\usepackage[flushleft]{threeparttable}
\usepackage{multirow}
\usepackage{afterpage}
\usepackage{hyperref}
\usepackage{natbib}
\usepackage{graphicx}
\hypersetup{
    colorlinks=true,
    filecolor=magenta,      
    citecolor=blue,
}

\modulolinenumbers[5]

\def\argmin{\mathop{\rm arg\,min}\limits}
\def\argmax{\mathop{\rm arg\,max}\limits}

\newcommand{\R}{\mathbb{R}}

\def\argmin{\mathop{\rm arg\,min}\limits}
\def\argmax{\mathop{\rm arg\,max}\limits}

\begin{document}

\title{An empirical comparison and characterisation of nine popular clustering methods}

\author{Christian Hennig,\\
Dipartimento di Scienze Statistiche ``Paolo Fortunati''\\
       Universita di Bologna\\
       Bologna, Via delle belle Arti, 41, 40126, Italy\\
christian.hennig@unibo.it}

\maketitle

\begin{abstract}
Nine popular clustering methods are applied to 42 real data sets. The aim is 
to give a detailed characterisation of the methods by means of several cluster
validation indexes that measure various individual aspects of the resulting 
clusters such as small within-cluster distances, separation of clusters, 
closeness to a Gaussian distribution etc. as introduced in \cite{Hennig19}.
30 of the data sets come with a ``true'' clustering. On these data sets 
the similarity of the clusterings from the nine methods to the ``true'' 
clusterings is explored. Furthermore, a mixed effects regression relates
the observable individual aspects of the clusters to the similarity with 
the ``true''
clusterings, which in real clustering problems is unobservable. 
The study gives new insight not only into the ability of the
methods to discover ``true'' clusterings, but also into properties of 
clusterings that can be expected from the methods, which is crucial
for the choice of a method in a real situation without a given ``true'' 
clustering.

\end{abstract}

\section{Introduction}
\label{sintro}
This work compares cluster analysis methods empirically on 42 real data sets.
30 of these data sets come with a given ``true'' classification. The principal
aim is to explore how different
clustering methods produce solutions with different data analytic
characteristics, which can help a user choosing an appropriate method for
the research question of interest. This does not require the knowledge of a 
``true'' clustering. The performance
of the methods regarding recovery of the ``truth'' is reported, but
is not the main focus.
 
Cluster analysis plays a central role in modern data analysis and is applied in 
almost every field where data arise, be it finance, marketing, genetics, 
medicine, psychology, archaeology, social and political science, chemistry, 
engineering, or machine learning. Cluster analysis can have well-defined 
research aims such as species delimitation in biology, or be applied in a
rather exploratory manner to learn about potentially informative structure in 
a data set, for example when clustering the districts of a city.  
New cluster analysis methods are regularly
developed, often for new data formats, but also to fix apparent defects
of already existing methods. One reason for this is that cluster analysis is 
difficult, and all methods, or at least those with which enough experience has
been collected, are known to ``fail'' in certain, even fairly regular and 
non-pathological, situations, where ``failing'' is often taken to mean 
that a certain pre-specified ``true'' clustering in data is not recovered. 

A key problem with clustering is that there is no unique and generally accepted
definition of what constitutes a cluster. This is not an accident, but rather 
part of the nature of the clustering problem. In real applications there can be
different requirements for a ``good'' clustering, and different clusterings
can qualify as ``true'' on the same data set. For example, crabs can be 
classified according to species, or as male or female; paintings can be 
classified according to style of the painter or according to the motif; a data
set of customers of a company may not show any clusters that are clearly 
separated from each other, but may be very heterogeneous, and 
the company may be interested in having homogeneous subgroups of customers in 
order to better target their campaigns, but the data set may allow for 
different groupings of similar quality; in many situations with given ``true''
classes, such as companies that go bankrupt in a given period vs. those that
do not, there is no guarantee that these ``true'' classes correspond to patterns
in the data that can be found at all. One could even argue that in a data set
that comes with a supposedly ``true'' grouping a clustering that does {\it not}
coincide with that grouping is of more scientific interest than reproducing 
what is already known.

Rather than being generally ``better'' or ``worse'', different cluster 
analysis methods can be seen as each coming with their own implicit 
definition of what a cluster is, and when cluster analysis is to be applied,
the researchers have to decide which cluster concepts are appropriate for the 
application at hand. Cluster analysis can have various aims, and these aims 
can be in conflict with each other. For example, clusters that are well
separated by clear density gaps may involve quite large 
within-cluster distances, which may be tolerable in some applications but 
unacceptable in others. Clusters that can be well represented by cluster 
centroids may be different from those that correspond to separable Gaussian
distributions with potentially different covariance matrices, which in some
applications are interpreted as meaningfully different data subsets.
See \cite{AckBDLok10,LuWiGu12,Hennig15,Hennig15handbook} for the underlying
``philosophy'' of clustering.
 
The starting point of this work is the collection of cluster validation indexes
presented in \cite{Hennig19}. These are indexes defined in order to provide 
a multivariate characterisation of a clustering, individually measuring aspects
such as between-cluster separation, within-cluster homogeneity, or 
representation of the overall dissimilarity structure by the clustering. They 
are applied here in order to give general information about how the 
characteristics of clusterings depend
on the clustering method. 

Many cluster validation indexes have been proposed in the literature, often in
order to pick an optimal clustering in a given situation, e.g., by comparing 
different numbers of clusters, see \cite{HaVaHe15} for an overview. Most of 
them (such as the Average Silhouette Width, \cite{KauRou90}) attempt to assess
the quality of a clustering overall by defining
a compromise of various aspects, particularly within-cluster homogeneity and 
between-cluster separation. 
Following \cite{Hennig19} and \cite{AkhHen20}, the present work deviates 
from this approach by 
keeping different aspects separate in order to inform the user in a more 
detailed way what a given clustering achieves. 

A number of benchmark studies for cluster analysis have already been published.
Most of them focus on evaluating the quality of clusterings by comparing them
to given ``true'' clusterings. This has been done for artificially generated
data (e.g., \cite{Mil80,BruSte07,SteBru11,SaDoDo13,RCCBACR19}; see \cite{Mil96}
for an overview of earlier work), for real data,
mostly focusing on specific application areas or types of data
(e.g., \cite{dCdLS08,KoPeWa14,BouHat17,LSWZYLD19}), or 
a mixed collection of real and
artificial data, sometimes generating artificial data from models closely 
derived from a real application 
(e.g., \cite{MeiHec01,MauBan02,DiBaWiHoMo04,AGMPP13,JLR20}). An exception is 
\cite{JTLB04}, where different clustering methods were mapped according to
the similarity of their clusterings on various data sets (something similar
is done here, see Section \ref{scharacterisation}). 
\cite{AndHen14} contrasted recovery
of a ``true'' classification in artificial data sets with the requirement of
having homogeneous clusters. 

All of these studies attempt to provide a neutral
comparison of clustering methods, which is to be distinguished from the large
number of studies, using real and artificial data, that have been carried out 
by method developers in order to demonstrate that their newly proposed method 
compares favourably with existing methods. Due to selection effects, the results
of such work, although of some value in their own right, cannot be taken as 
objective indicators of the quality of methods 
(\cite{BoLaEu13,Hennig18}).
The study presented here is meant to be neutral; I have not been 
involved in the development of any of the compared methods, and have no specific
interest to portray any of them as particularly good or bad. 
Note that ``true'' 
neutrality can never be secured and is probably never given; for example, 
I have been active promoting my own ``philosophy'' 
of clustering (e.g., \cite{Hennig15handbook}) and may be suspected to favour 
results that are in line with the idea that the success of clustering methods 
strongly depends on the application; however n
No selections have been made depending on results (\cite{Boulesteix15}); 
the 42 data sets 
from which results are 
reported are all that were involved. 

Section \ref{sdesign} explains the design of the study, i.e., the 
clustering methods, the data sets, and the validation indexes. Section 
\ref{sresults} presents the results, starting with the characterisation of the
methods in terms of the internal indexes, then results regarding the recovery
of the ``true'' clusters, and ultimately connecting ``true'' cluster recovery
with the characteristics of the clustering solutions using a mixed effects 
regression model. A discussion concludes the paper. 

\section{Study design}
\label{sdesign}
For the study design, recommendations for benchmark studies as given, e.g., in
\cite{Boulesteix15,VBDDGHLS18} have been taken into account. One important issue
is a definition of the scope of the study.
There is an enormous amount of clustering methods, and clustering is applied to 
data of very different formats. It is not even remotely possible to cover 
everything that could potentially be of interest. Therefore the present study
constrains its scope in the following way:
\begin{itemize}
\item Only clustering methods for $2\le p$-dimensional Euclidean data that 
can be treated as continuous are used. Methods that work with dissimilarities
are run using the Euclidean distance.
\item Accordingly, data sets contain numerical variables only.
Some data sets include discrete variables, which are treated as admissible 
for the study if they carry numerical information and take at least three 
different values (variables taking a small number of values, particularly 
three or four, are very rare in the study).
\item The number of clusters is always treated as fixed. Only methods that allow
to fix the number of clusters are used; methods to estimate the
number of clusters are not involved. For data sets with a given 
``true'' clustering, the corresponding number of clusters was taken. For data
sets without such information, a number of clusters was chosen subjectively 
considering data visualisation and, where possible, subject matter information.
\item The included clustering methods were required to have an 
R-implementation that can be used in a 
default way without additional tuning in order to allow for a comparison that
is not influenced by different tuning flexibilities. 
\item No statistical structure (such as time series or regression clustering) 
is taken into account, and neither is any automatic dimension reduction involved
as part of any method. All data is treated as plain $p$-dimensional Euclidean.
\item Methods are only admissible for the study if they always produce crisp
partitions. Every observation always is 
classified (also in the given ``true'' clusterings) 
to belong to one and only one cluster. 
\end{itemize}
\subsection{Clustering methods}
\label{sclustering}
The involved clustering methods are all well established and widely used, as 
far as my knowledge goes. 
They represent the major classes of clustering methods listed in 
\cite{HenMei15} with the exception of density-based clustering, which
was excluded because standard density-based methods such as DBSCAN 
(\cite{EsKrSaXu96}) do not accept the number of clusters as input and often 
do not produce partitions. Another popular method that was not involved was
Ward's method, as this is based on the same objective function as $K$-means and
can be seen as just another technique to optimise this function locally 
(\cite{EvLaLeSt11}). On the other hand, including mixtures of t- and skew 
t-distributions means that mixture model-based clustering is strongly 
represented. The motivation for this is that the other included methods
are not meant to fit distributional shapes including outliers and skewness,
which may be widespread in practice; alternatives would be methods that have
the ability to not include observations classified as ``outliers'' in any 
cluster, but this is beyond the scope of the present study.
Here are the included methods. 
\begin{description}
\item[K-means] as implemented in the R-function \verb|kmeans| using the
algorithm by \cite{HarWon79}. 
\item[Partitioning Around Medoids (clara)] (\cite{KauRou90}) as implemented 
in the 
R-function \verb|claraCBI| (therefore abbreviated ``clara'' in the results)
in R-package \verb|fpc| (\cite{Hennig20}), which calls function
\verb|pam| in R-package \verb|cluster| (\cite{MRSHH19}) 
using (unsquared) Euclidean distances.  
\item[Gaussian mixture model (mclust)] fitted by Maximum Likelihood using the 
EM-algorithm, where the best of various covariance matrix models is chosen
by the Bayesian Information Criterion (BIC) (\cite{FraRaf02}) as implemented
in the R-function \verb|mclustBIC| in R-package \verb|mclust| 
(\cite{ScFoMuRa16}). 
\item[Mixture of skew t-distributions (emskewt)] fitted by 
Maximum Likelihood using the EM-algorithm (\cite{LeeMcL13}), including fully
flexible estimation of the degrees of freedom and the shape matrix, as 
implemented  in the function \verb|EmSkew| with parameter \verb|distr="mst"| 
in the R-package \verb|EMMIXskew|
(\cite{WaNgMc18}). 
\item[Mixture of t-distributions (teigen)] fitted by Maximum Likelihood 
using the 
EM-algorithm (\cite{McLPee00}), where the best of various covariance matrix 
models is chosen
by the BIC (\cite{AndMcN12}) as implemented in the R-function \verb|teigen|
in R-package \verb|teigen|
(\cite{AnWiBoMc18}).
\item[Single linkage hierarchical clustering] as implemented in the R-function
\verb|hclust| and the dendrogram cut at the required number of clusters
to produce a partition, as is done also for the other hierarchical methods. 
See \cite{EvLaLeSt11} for an explanation and 
historical references for all involved hierarchical methods.
\item[Average linkage hierarchical clustering] as implemented in the R-function
\verb|hclust|.
\item[Complete linkage hierarchical clustering] as implemented in the R-function
\verb|hclust|.
\item[Spectral clustering] (\cite{NgJoWe01}) as implemented in the R-function
\verb|specc| in R-package \verb|kernlab| (\cite{KaSmHoZe04}).
\end{description}
The functions were mostly run using the default settings. In some cases,
e.g., \verb|hclust|, parameters had to be provided in order to 
determine which exact method was used. Some amendments were required. In 
particular, all methods were run in such a way that they would always
deliver a valid partition as a result. See Appendix 
A1 for more computational detail.
\subsection{Data sets}
The data sets used in this study are a convenience sample, collected from
mostly well known benchmark data sets in widespread use together with 
some data sets that I have come across in my work. 21 data sets are from
the UCI repository (\cite{UCI17}), further ones are from Kaggle,
\verb|www.openml.org|, example data sets of R-packages, 
open data accompanying books and research papers,
and some were
collected by myself or provided to me by collaborators and advisory
clients with permission to use them. Details about the data sets are given 
in Appendix A2.

There were 
some criteria on top of those stated above according to which data sets 
have been selected, which define the scope of the study. There was a target
number of collecting at least 30 data sets with and at least 10 data sets
without given ``true'' classes; ultimately there are 30 data sets with and 12
data sets without true classes.
The aim was to cover a large range of application areas, although due to the
availability of data sets, this has not been perfectly achieved. 17 of the 
data sets come from the related areas of biology, genetics, medicine, and 
chemistry. 
Eight are from the social sciences, two from finance, eight can be classified 
as engineering including typical pattern recognition tasks, the 
remaining seven data sets come from miscellaneous areas.
  
As some of the clustering methods cannot handle data with a smaller number of 
observations $n$ than the number of variables $p$ within clusters, all data sets
have $p$ substantially smaller than $n$. The calibration of validation indexes 
requires repeated computations based on $n\times n$ distance matrices 
(see Section \ref{sinternal}), for this reason the biggest data set has 
$n=4601$, and generally data sets with $n<3000$ were preferred. The maximum 
$p$ is 72. $p=1$ is excluded, as it could not be handled by two methods.
The maximum number of ``true'' clusters $K$ is 100.
The aim was to achieve a fairly even
representation of $p$ and $K$ up to 10 and a number of instances for these
values larger than 10, although there are apparently far more data sets in
benchmark use with $k=2$ than with larger $K$. 
Data sets 
without missing values were preferred, but some data sets with a very small 
number of missing values were admitted. In these cases mean imputation was used.
Tables \ref{tn}, \ref{tp}, and \ref{tk} show the distributions of $n$, $p$, and
$K$, respectively, over the data sets.

The variables were scaled to mean 0 and variance 1 before clustering, except 
for data sets in which the variables have compatible units of measurement and
there seems to be a subject matter justification to make their impact for
clustering proportional to the standard deviation. See Appendix 
A2 for details on the preprocessing for some data sets. 

\begin{table}[tb]
\caption{Numbers of observations for the 42 data sets.} 
\begin{tabular}{|l|r|}
\hline
Observations & Number of data sets\\
\hline
$n\le 100$ & 5 \\
$100<n\le 200$ & 6\\ 
$200<n\le 300$ & 8\\
$300<n\le 500$ & 5\\
$500\le n<1000$ & 7\\
$1000\le n<2000$ & 6\\
$n>2000$ & 5\\
\hline
\end{tabular}
\label{tn}
\end{table}

\begin{table}[tb]
\caption{Numbers of variables for the 42 data sets}
\begin{tabular}{|l|r|}
\hline
Variables & Number of data sets\\
\hline
$p=2$ & 2\\
$p=4$ & 5\\ 
$p=5$ & 5\\
$6\le p\le 8$ & 6\\
$9\le p \le 11$ & 11\\
$12\le p \le 20$ & 6\\
$21\le p \le 50$ & 4\\
$p > 50$ & 3\\
\hline
\end{tabular}
\label{tp}
\end{table}

\begin{table}[tb]
\caption{Numbers of clusters for the 30 data sets with given ``true'' 
clusterings, and for the 12 data sets without ``true'' clusterings, as chosen
by the author.}
\begin{tabular}{|l|r|r|}
\hline
Number of clusters & With ``true'' clustering & Without ``true'' clustering\\
\hline
$k=2$ & 8 & 1\\
$k=3$ & 3 & 3\\ 
$k=4$ & 3 & 1\\
$k=5$ & 2 & 6\\
$6\le k\le 7$ & 5 & 1\\
$8\le k \le 11$ & 6 & 0\\
$k>11$ & 3 & 0\\
\hline
\end{tabular}
\label{tk}
\end{table}

An issue with the ``representativity'' of these data sets for real clustering problems is that the availability of ``true'' clusterings constitutes a difference to the real unsupervised problems to which clustering is usually applied. This is an issue with almost all collections of data sets for benchmarking clustering algorithms. In particular, several such data sets have been constructed in order to have all clusters represented by the same number of observations. This is the case for eight of the 30 data sets with ``true'' clusterings used here (seven of these have exactly equal cluster sizes).
This is not possible for unsupervised problems in practice. Such data sets will favour methods that tend to produce clusters of about equal sizes.

\subsection{Internal validation indexes}
\label{sinternal}
Internal validation indexes are used here with the aim of measuring various 
aspects of a clustering that can be seen as desirable, depending on the specific
application. It is then investigated to what extent the different clustering
methods work well according to these aspects. \cite{Hennig15handbook} lists and
discusses a number of aspects that can be relevant. \cite{Hennig19} and
\cite{AkhHen20} formalised many of these aspects, partly using already existing
indexes, partly introducing new ones. Here the indexes used in the present study
are listed. For more background and discussion, including possible alternatives,
see \cite{Hennig19} and
\cite{AkhHen20}. The indexes attempt to formalise clustering aspects in a direct
intuitive manner, without making reference to specific models (unless it is of
interest whether data look like generated by a particular probability model, 
see below). 
The indexes as defined here do not allow comparison between
or aggregation over different data sets. In order to do this, they need to be 
calibrated, which is treated in Section \ref{scalibration}.

The data set is denoted as
${\cal D}=\{x_1,\ldots,x_n\}$. Here the observations 
$x_1,\ldots,x_n$ are assumed to be $\in\R^p$, and $d(x,y)$ is the Euclidean
distance between $x$ and $y$, although the indexes can be applied
to more general types of data and distances. A clustering is 
a set ${\cal C}=\{C_1,\ldots,C_K\}$ with 
$C_j\subseteq {\cal D},\ j=1,\ldots,K$. For $j=1,\ldots,K$, $n_j=|C_j|$ is 
the number of objects in $C_j$. Assume ${\cal C}$ 
to be a partition, e.g., $j\neq k \Rightarrow C_j \cap C_k=\emptyset$ and 
$\bigcup_{j=1}^K C_j={\cal D}$. Let 
$\gamma:\ \{1,\ldots,n\}\mapsto \{1,\ldots,K\}$ be the assignment function, 
i.e., $\gamma(i)=j \Leftrightarrow x_i\in C_j$.
\begin{description}
\item[Average within-cluster distances] (avewithin; aw; \cite{AkhHen20}). 
This index measures homogeneity in the
sense of small distances within clusters. Smaller values are better.
\begin{displaymath}
I_{avewithin}(\mathcal{C}) = \frac{1}{n} \sum_{k=1}^{K} \frac{1}{n_k-1}\sum_{x_{i} \neq x_{j} \in C_{k}} d(x_{i},x_{j}). 
\end{displaymath}	
\item[Representation of cluster members by centroids.] In some applications
cluster centroids are used in order to represent the clustered objects, and
an important aim is that this representation is good for all cluster members.
This is directly formalised by the objective functions of $K$-means (sum of
squared distances from the cluster mean) and Partitioning Around Medoids (sum
of distances from the cluster medoid). Both of these criteria have been used
as internal validation indexes in the present study, however results are
not presented, because over all results both of these turn out to have a 
correlation of larger than 0.95 with $I_{avewithin}$, so $I_{avewithin}$ can be 
taken to measure this clustering aspect as well.  
\item[Maximum diameter] (maxdiameter; md). In some applications there may be 
a stricter requirement that large distances within clusters cannot be tolerated,
rather than having only the distance average small. This can be formalised by
\begin{displaymath}
  I_{maxdiameter}(\mathcal{C})=\max_{C\in\mathcal{C};x_i,x_j\in C}d(x_i,x_j).
\end{displaymath}
Smaller values are better.
\item[Widest within-cluster gap] (widestgap; wg; \cite{Hennig19}). Another 
interpretation of cluster homogeneity is that there should not be 
different parts of the same cluster that are separated from each other. This 
can be formalised by
\begin{displaymath}
  I_{widestgap}({\cal C})=\max_{C\in {\cal C}, D, E:\ C=D\cup E}\min_{x\in D, y\in E}d(x,y).  
\end{displaymath}
Smaller values are better.
\item[Separation index] (sindex; si; \cite{Hennig19}). This index measures 
whether clusters are
separated in the sense that the closest distances between clusters are large.
For every object $x_{i} \in C_{k}$, $i = 1, \ldots, n$, $k \in {1, \ldots, K}$, let $d_{k:i} = \min_{x_{j} \notin C_{k}} d(x_{i},x_{j})$. Let $d_{k:(1)} \leq \ldots \leq d_{k:(n_{k})}$ be the values of $d_{k:i}$ for $x_{i} \in C_{k}$ ordered from the smallest to the largest, and let $[pn_{k}]$  be the largest integer $\leq pn_{k}$. $p$ is a parameter tuning what proportion of observations counts as
``close to the border'' of a cluster with another. Here, $p=0.1$.
Then, 
\begin{displaymath}
I_{sindex}(\mathcal{C};p) = \frac{1}{\sum_{k=1}^{K} [pn_{k}]} \sum_{k=1}^{K} \sum_{i=1}^{[pn_{k}]} d_{k:(i)}.  
\end{displaymath}
Larger values are better. 

Analogously to the maximum diameter, the minimum 
separation, i.e., the minimum distance between any two clusters may also be 
of interest. In the present study, this has a correlation of 0.93 with 
$I_{sindex}$, and results for the minimum separation are omitted for reasons
of redundancy.

\item[Pearson-version of Hubert's $\Gamma$] (pearsongamma; pg; \cite{HubSch76}).
This index measures to what extent the clustering corresponds or represents 
the distance structure in the data.  
the vector of pairwise dissimilarities
Let ${\bf d}={\rm vec}\left([d(x_i,x_j)]_{i<j}\right)$ be the vector of 
pairwise distances. Let ${\bf c}={\rm vec}\left([c_{ij}]_{i<j}\right)$, where $c_{ij}=1(\gamma(i)\neq\gamma(j))$, and $1(\bullet)$ denotes the indicator function, be a vector of ``clustering induced dissimilarities''. With $r$ denoting the sample Pearson correlation,
  \begin{displaymath}
    I_{Pearson\Gamma}({\cal C})=r({\bf d},{\bf c}).
  \end{displaymath}
Larger values are better. This is one version of a family of indexes introduced
in \cite{HubSch76}, sometimes referred to as ``Hubert's $\Gamma$''.
\item[Density mode index] (dmode; dm). An intuitive idea of a cluster 
is that it
is associated with a density mode, and that 
the density goes down toward the cluster 
border. This is formalised by the ``dmode'' index. 
It is based on a simple kernel density
estimator $h$ that assigns a density value $h(x)$ to every observation. 
Let $q_{d,p}$ be the $p$-quantile of the vector of dissimilarities 
${\bf d}$, e.g., for $p=0.1$, the 10\% smallest dissimilarities are 
$\le q_{d,0,1}$. Define the kernel and density as 
\begin{displaymath}
\kappa(d)=\left(1-\frac{1}{q_{d,p}}d\right)1(d\le q_{d,p}),\qquad
h(x)=\sum_{i=1}^n \kappa(d(x,x_i)).
\end{displaymath}
The following algorithm constructs a sequence of neighbouring
observations from the mode in such a way that the density should always go 
down, and penalties are incurred if the density goes up.
It also constructs a set $T$ that collects 
information about high dissimilarities between high density observations used
below. $I_{densdec}$ collects the penalties.
\begin{description}
\item[Initialisation] $I_{d1}=0$, $T=\emptyset$. For $j=1,\ldots,K$: 
\item[Step 1] $S_j=\{x\}$, where $x=\argmax_{y \in C_j}h(y)$. 
\item[Step 2] Let $R_j=C_j\setminus S_j$. If $R_j=\emptyset$: $j=j+1$,
if $j\le K$ go to Step 1, if $j+K=1$ then go to Step 5. Otherwise:
\item[Step 3] Find $(x,y)=\argmin_{(z_1,z_2): z_1\in R_j, z_2\in S_j}d(z_1,z_2)$.
$S_j=S_j\cup\{x\}$, $T=T\cup \{\max_{z\in R_j}h(z)d(x,y)\}$. 
\item[Step 4] If $h(x)>h(y):\ I_{d1}=I_{d1}+(h(x)-h(y))^2$, back to Step 2.
\item[Step 5] $I_{densdec}({\cal C})=\sqrt{\frac{I_{d1}}{n}}.$   
\end{description}
It is possible that there
is a large gap between two observations with high density, which does not
incur penalties in $I_{densdec}$ if there are no low-density observations 
in between. This can be picked up by 
\begin{displaymath}
  I_{highdgap}({\cal C})=\max T.  
\end{displaymath}
These two indexes, which are both better for smaller values, 
were defined in \cite{Hennig19}, but they can be 
seen as contributing to the measurement of the same aspect, with 
$I_{highdgap}$ just adding information missed by $I_{densdec}$. An 
aggregate version, which is used here, can be defined as
\begin{displaymath}
  I_{dmode}({\cal C})=0.75 I_{densdec}^*({\cal C})+0.25 I_{highdgap}^*({\cal C}),
\end{displaymath}
where $I_{densdec}^*$ and $I_{highdgap}^*$ are suitably calibrated versions of
$I_{densdec}$, $I_{highdgap}$, respectively, see Section \ref{scalibration}. 
The weights 0.75 and 0.25 in the definition of $I_{dmode}$ can be 
interpreted as the relative impact of the two sub-indexes.
\item[Cluster boundaries cutting through density valleys] (denscut; dc;
\cite{Hennig19}). A complementary aspect of the idea that clusters are 
associated with high density regions is that cluster boundaries should run
through density valleys rather than density mountains. The ``denscut''-index 
penalises a high contribution of points from different 
clusters to the density values in a cluster (measured by $h_o$ below).
\begin{displaymath}
\mbox{For }x_i,\ i=1,\ldots,n:\  h_o(x_i)=\sum_{k=1}^n \kappa(d(x_i,x_k))
1(\gamma(k)\neq\gamma(i)).
\end{displaymath}
A penalty is incurred if for observations with a large density $h(x)$
there is a large contribution $h_o(x)$ to that density from other clusters:
\begin{displaymath}
  I_{denscut}({\cal C})=\frac{1}{n}\sum_{j=1}^K\sum_{x\in C_j} h(x)h_o(x).
\end{displaymath}
Smaller values are better.
\item[Entropy] (en; \cite{Shannon48}). 
Although not normally listed as primary aim of clustering, 
in many applications very small clusters are not very useful, and cluster sizes should optimally be close to uniform. This is measured by the well known entropy: 
	\begin{displaymath}
		I_{entropy}(\mathcal{C}) = - \sum_{k=1}^{K} \frac{n_{k}}{n} \log(\frac{n_{k}}{n}).
	\end{displaymath}
Large values are good. 
\item[Gaussianity of clusters] (kdnorm; nor; \cite{CorHen16}). Due to the 
Central Limit Theorem and a widespread belief that the Gaussian distribution 
approximates many real random processes, it may be of 
interest in its own right to have clusters that are approximately Gaussian.
The index $I_{kdnorm}$ is defined, following  \cite{CorHen16}, as the 
Kolmogorov distance between the empirical distribution of within-cluster
Mahalanobis distances to the cluster means, and a $\chi^2_p$-distribution, 
which is the distribution of Mahalanobis distances in 
perfectly Gaussian clusters. 
\item[Coefficient of variation of distances to within-cluster neighbours]
(cvnnd; cvn; \cite{Hennig19}). Another within-cluster distributional shape 
of potential interest is uniformity, where clusters are characterised by
a uniform within-cluster density level. 
This can be characterised by the coefficient of variation (CV) of 
the dissimilarities to the $k$th nearest 
within-cluster neighbour $d^k_w(x)$ ($k=2$ is used here). 
Define for $j=1,\ldots,k$, assuming $n_j>k$:
\begin{displaymath}
  m(C_j;k)=\frac{1}{n_j}\sum_{x\in C_j}d^k_w(x),\qquad  
{\rm CV}(C_j)=\frac{\sqrt{\frac{1}{n_j-1}\sum_{x\in C_j}(d^k_w(x)-m(C_j;k))^2}}
{m(C_j;k)}.
\end{displaymath}
Using this,
\begin{displaymath}
  I_{cvdens}({\cal C})=\frac{\sum_{j=1}^K n_j {\rm CV}(C_j)1(n_j>k)}
{\sum_{j=1}^K  n_j1(n_j>k)}.
\end{displaymath}
Smaller values are better.
\item[Average Silhouette Width] (asw; \cite{KauRou90}). This is a popular 
internal validation index that deviates somewhat 
from the ``philosophy'' behind the
collection of indexes presented here, because it attempts to balance 
two aspects of cluster quality, namely homogeneity and separation. It has 
been included in the study anyway, because it also uses an intuitive direct 
formalisation of clustering characteristics of interest. 
For $i = 1, \ldots, n$, define the ``silhouette width'' 
\begin{displaymath}
s_{i}=\frac{b_{i}-a_{i}}{\max{\left\{a_{i}, b_{i}\right\}}} \in [-1,1],
\end{displaymath}
where		
\begin{displaymath}
a_{i}=\frac{1}{n_{l_i}-1} \sum_{x_{j} \in C_{l_i}} d(x_{i}, x_{j}),\
b_{i}=\min_{h \neq l_i} \frac{1}{n_{h}} \sum_{x_{j} \in C_{h}} d(x_{i}, x_{j}). 
\end{displaymath}
The Average Silhouette Width is then defined as
\begin{displaymath}
I_{asw}(\mathcal{C}) = \frac{1}{n} \sum_{i=1}^{n} s_{i}.
 \end{displaymath}
\end{description}

\subsection{Calibrating the indexes}
\label{scalibration} 
For aggregating the indexes introduced in Section \ref{sinternal} over different
data sets and to compare the performance of a clustering method over the 
indexes in order to characterise it, it is necessary to calibrate the values of 
the indexes, so that they become comparable. This is done as in 
\cite{Hennig19,AkhHen20}. The idea is to generate a large number $m$ 
of ``random clusterings'' $\mathcal{C}_{R1},\ldots,\mathcal{C}_{Rm}$ 
on the data. Denote the clusterings of the $q=9$ methods from Section 
\ref{sclustering} by ${\mathcal C}_1,\ldots,\mathcal{C}_q$. For a given 
data set ${\cal D}$ and index $I$, first change $I$ to $-I$ in case that 
smaller values are better according to the original definition of $I$, so that
for all calibrated 
indexes larger values are better. Then use these clusterings to
standardise $I$:
\begin{eqnarray*}
m(I,{\cal D})&=&\frac{1}{m+q}\left(\sum_{i=1}^m I(\mathcal{C}_{Ri})+ \sum_{i=1}^q I(\mathcal{C}_{i})\right),\\ 
s^2(I,{\cal D})&=& \frac{1}{m+q-1}\left(\sum_{i=1}^m \left[I(\mathcal{C}_{Ri})-
m(I,{\cal D})\right]^2+ \sum_{i=1}^q \left[I(\mathcal{C}_{i})-m(I,{\cal D})\right]^2\right),\\
I^*(\mathcal{C}_{i})&=&\frac{I(\mathcal{C}_i)-m(I,{\cal D})}{s(I,{\cal D})},\ 
i=1,\ldots,q.
\end{eqnarray*}
$I^*$ is therefore scaled so that its values can be interpreted as expressing 
the quality (larger is better) compared to what the collection of clusterings
$\mathcal{C}_{R1},\ldots,\mathcal{C}_{Rm},{\mathcal C}_1,\ldots,\mathcal{C}_q$
achieves on the same data set. The approach depends on the definition of the
random clusterings. These should generate enough random variation in order to 
work as a tool for calibration, but they also need to be reasonable 
as clusterings, because if all random clusterings
are several standard deviations away from the ``proper'' clusterings, the 
exact distance may not be very meaningful.
They also need to be fast to generate, as many of them will be required in order
to calibrate index values of every single data set.

Four different algorithms are used for generating the random clusterings, for
detains see \cite{AkhHen20}. For clusterings
with $K$ clusters, these are:
\begin{description}
\item[Random $K$-centroids:] Draw $K$ observations from ${\cal D}$. Assign every
observation to the nearest centroid.
\item[Random nearest neighbour:] Draw $K$ observations as starting points for 
the $K$ clusters. At every stage, of the observations that are not yet 
clustered, assign the observation $x$ 
to the cluster of its nearest already clustered 
neighbour, where $x$ is the observation 
that has the smallest distance to this neighbour.
\item[Random farthest neighbour:] As random nearest neighbour, but
$x$ is the observation 
that has the smallest distance to the minimum farthest cluster member.
%
\item[Random average distances:] As random nearest neighbour, but
$x$ is the observation 
that has the smallest average distance to the closest cluster.
%
\end{description}
Experience shows that these methods generate a range of clusterings that
have sufficient variation in characteristics and are mostly reasonably
close to the proper clustering methods (as can be seen in 
\cite{AkhHen20} as 
well as from the results of the present study).
Here, 50 random clusterings from each algorithm are generated, i.e., $m=200$.
All results in Section \ref{sresults} are given in terms of calibrated indexes
$I^*$.
 
\subsection{External validation indexes}
``Truth'' recovery is measured by external validation indexes that quantify
the similarity between two clusterings on the same data, here the ``true'' one
and a clustering generated by one of the clustering methods.

The probably most popular external validation index is the Adjusted Rand Index
(ARI; \cite{HubAra85}). This index is based on the relative number of pairs of 
points that are in the same cluster in both clusterings or in different clusters
in both clusterings, adjusted for the number of clusters and the cluster
sizes in such a way that its expected value under random cluster labels with the
same number and sizes of clusters is 0. The maximum value is 1 for perfect 
agreement. Values can be negative, but already a value of 0 can be interpreted 
as indicating that the two clusterings have nothing to do with each other. 

In some work, the ARI has been criticised, often in the framework of an 
axiomatic approach where it can be 
shown that it violates some axioms taken to be desirable, e.g., 
\cite{Meila07,AGAV09}. Alternative indexes have been proposed
that fulfill the presented axioms. \cite{Meila07} introduced the Variation of
Information (VI), which is a proper metric between partitions. This
means that, as opposed to the ARI, smaller values are better. In
Section \ref{sresults}, 
the negative VI is considered so that for all considered indexes larger values
are better. The VI is defined by comparing the entropies of the two 
clusterings with the so-called ``mutual information'', which is based on the 
entropy of the intersections between two clusters from the two different 
clusterings. If the two clusterings are the same, the entropy of the
intersections between clusters is the same as the entropy of the original 
clusterings, meaning that the VI is zero, its minimum value.

\cite{AGAV09} show their axioms for an index called BCubed first proposed 
in \cite{BagBal98}. This index is based on observation-wise concepts of 
``precision'' and ``recall'', i.e., what percentage of observations in the same
cluster are from the same ``true'' class, and what percentage of observations 
in a different cluster is ``truly'' different. It takes values between 0 and 1,
1 corresponding to a perfect agreement. See \cite{Meila15} for further 
discussion and some more alternatives.

\section{Results}
\label{sresults}
Three issues are addressed: 
\begin{itemize}
\item How can the clusters produced by the methods be characterised in terms
of the external validation indexes?
\item How do the methods perform regarding the recovery of the ``true'' clusterings?
\item Can the recovery of the ``true'' clusterings be related to the internal validation indexes?
\end{itemize}
\subsection{Characterisation of the methods in terms of the internal indexes}
\label{scharacterisation}
The methods can be characterised by the distribution of values of the calibrated
internal validation indexes, highlighting the dominating features of the 
clusterings that they produce. In order to do this, parallel coordinate plots 
will be used that show the full results including how results belonging to the
same data set depend on each other. 

I decided against running null hypothesis tests due to issues of multiple 
testing and model assumptions; the plots allow a good assessment of to what 
extent differences between methods are meaningful, dominated by random 
variation, or borderline.
Although the values of the calibrated indexes can be
compared over indexes as relative to the ensemble of clusterings from the 
methods and random, what is shown are images that compare the different 
clustering methods for each index, as the comparison of the clustering 
methods gives information additional to the performance relative to the random
clusterings.
 
\begin{figure}
\begin{center}
  \includegraphics[width=0.48\textwidth]{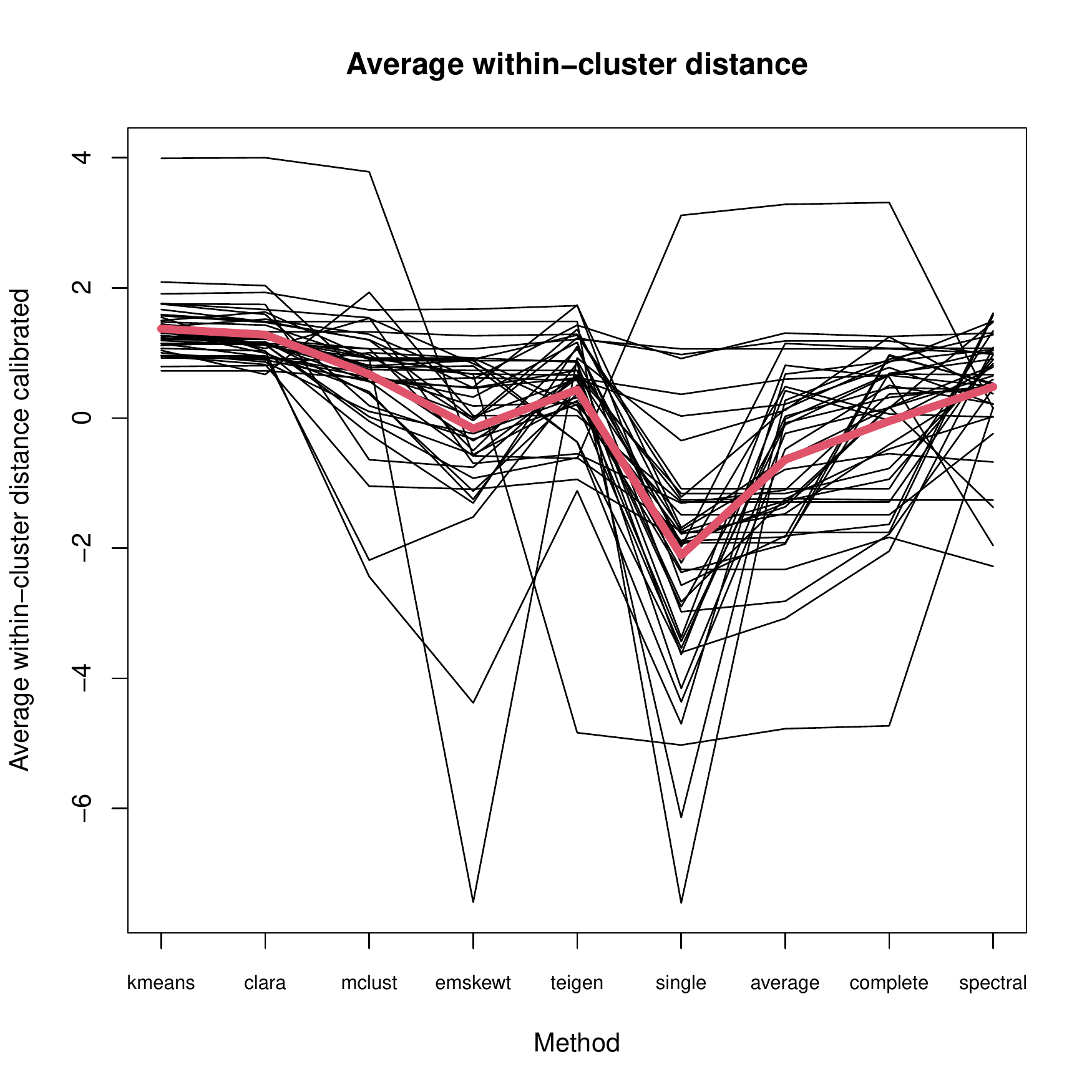}
  \includegraphics[width=0.48\textwidth]{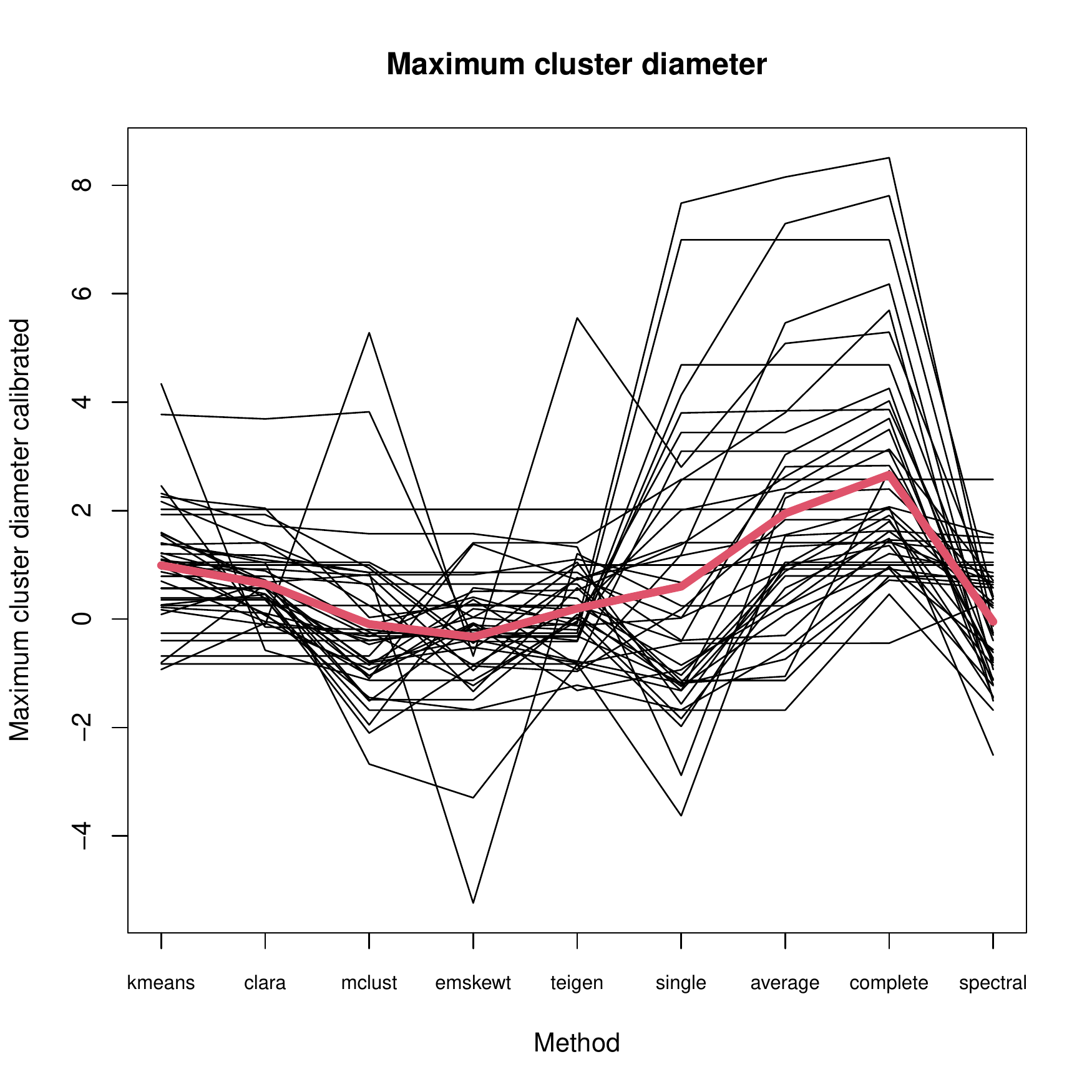}
\end{center}
\caption{Calibrated values of $I_{avewithin}^*$ and $I_{maxdiameter}^*$. Values belonging to the same data set are connected by lines. The thick red line gives the
average values.}
\label{fawmd}       
\end{figure}

\begin{description}
\item[Average within-cluster distances] (left side of Figure \ref{fawmd}):
The two centroid-based methods $K$-means and clara 
achieve the best results. The Gaussian and $t$-mixture are
about at the same level as spectral clustering; complete linkage and 
the mixture of skew $t$-distributions are worse. 
Average linkage is behind
these, and single linkage is the worst by some distance. 

Results regarding representation of the data by centroids are not shown and 
look largely the same. The only additional distinctive
feature is that $K$-means is better than clara looking at squared Euclidean 
distances to the centroid, whereas clara is better for unsquared distances.
This was to be expected, as it corresponds to what $K$-means, 
clara, respectively, attempt to optimise.

\item[Maximum diameter] (right side of Figure \ref{fawmd}): Unsurprisingly,
complete linkage is best; at each step it merges clusters so that the maximum 
diameter is the smallest possible, although it is not optimal for every single
data set (the hierarchical scheme will not normally produce a global optimum).
Average linkage is second best, followed by $K$-means, clara, and single 
linkage, which somewhat surprisingly 
avoids large distances within clusters 
more than spectral clustering and the three mixture models. Another potential 
surprise is that the Gaussian mixture does not do better than the $t$-mixture 
in this respect; a flexible covariance matrix can occasionally 
allow for very large within-cluster distances. 

\begin{figure}
\begin{center}
  \includegraphics[width=0.48\textwidth]{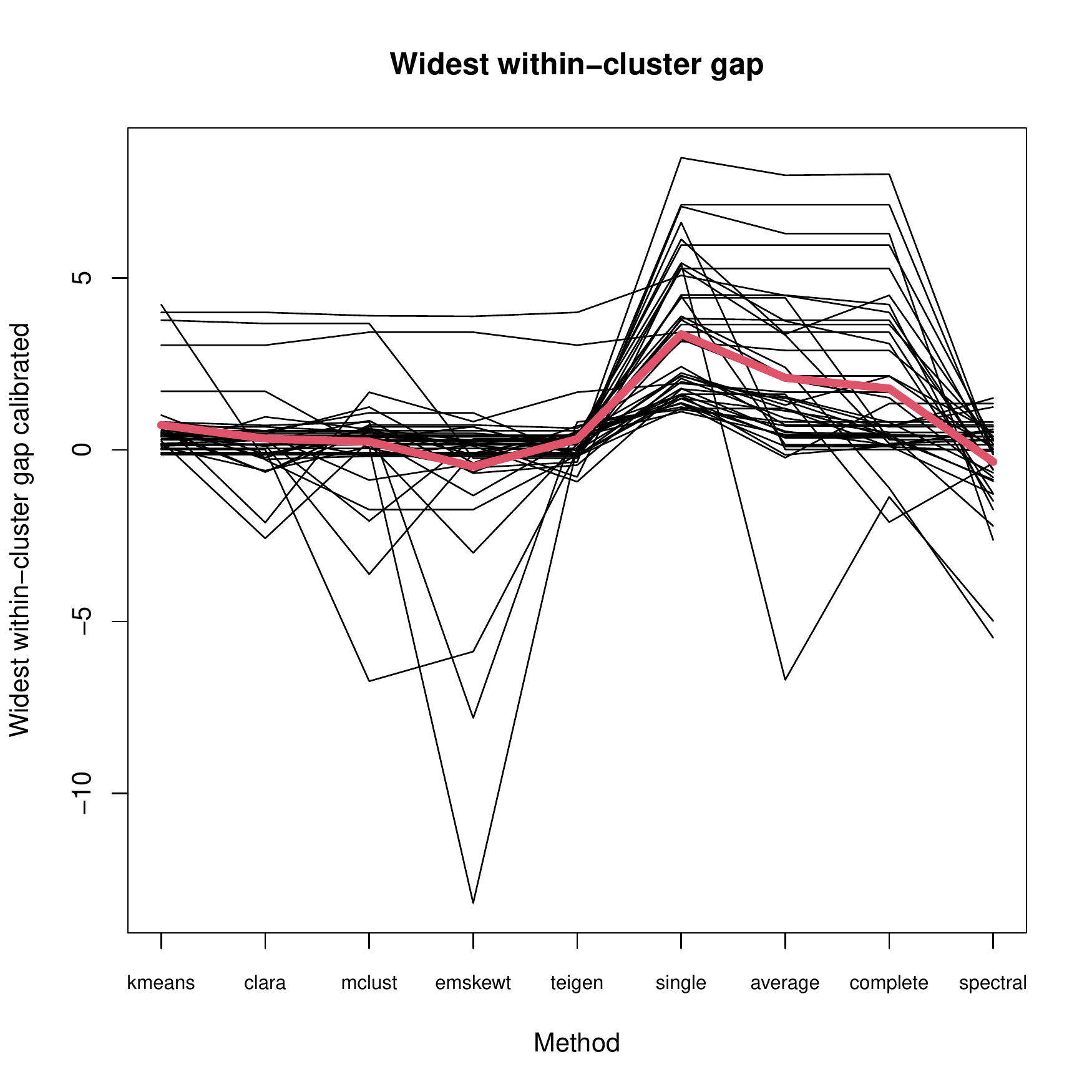}
  \includegraphics[width=0.48\textwidth]{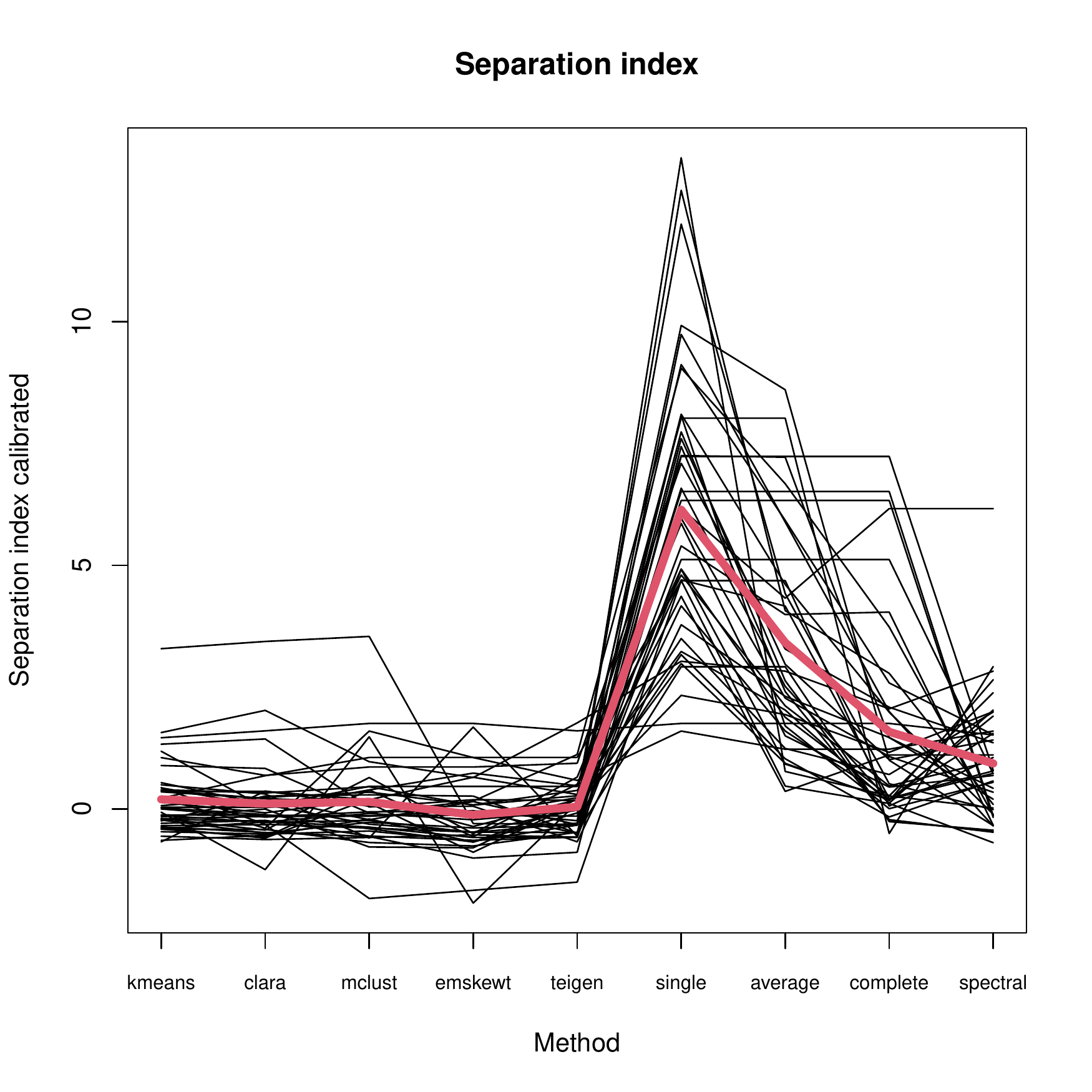}
\end{center}
\caption{Calibrated values of $I_{widestgap}^*$ and $I_{sindex}^*$. Values belonging to the same data set are connected by lines. The thick red line gives the
average values.}
\label{fwgsi}       
\end{figure}

\item[Widest within-cluster gap] (left side of Figure \ref{fwgsi}): The three
linkage methods are best at avoiding large within-cluster gaps, with single 
linkage in the first place, which will not join sets between which there is a 
large gap. The two centroid-based 
methods follow, however differences between them, the three mixture models, and
spectral clustering look small compared to the variance, and dominated by 
outliers. The skew $t$-mixture produces very large within-cluster gaps for a 
number of data sets. With strong skewness there can be large 
distances in a tail of a cluster.
\item[Separation index] (right side of Figure \ref{fwgsi}): Single linkage 
achieves the best results here. Its clustering process keep separated subsets 
in distinct clusters (often one-point clusters with strongly separated 
outliers). The two other linkage methods
follow. Complete linkage is sometimes portrayed as totally
prioritising within-cluster homogeneity over separation, but in fact regarding
separation it does better than spectral clustering, which is still a bit better
than the centroid-based and the mixture models, between 
which differences look insignificant.

\begin{figure}
\begin{center}
  \includegraphics[width=0.48\textwidth]{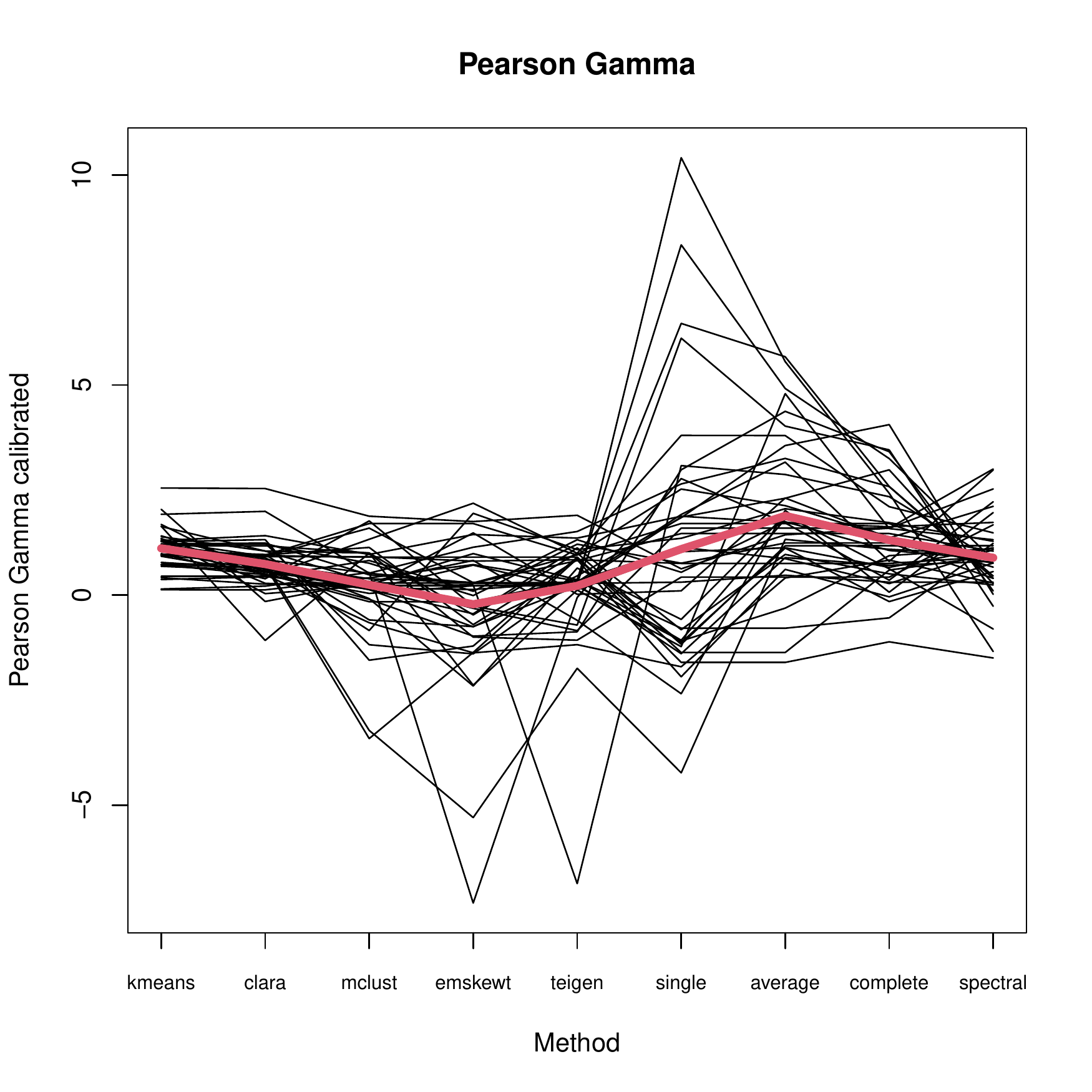}
  \includegraphics[width=0.48\textwidth]{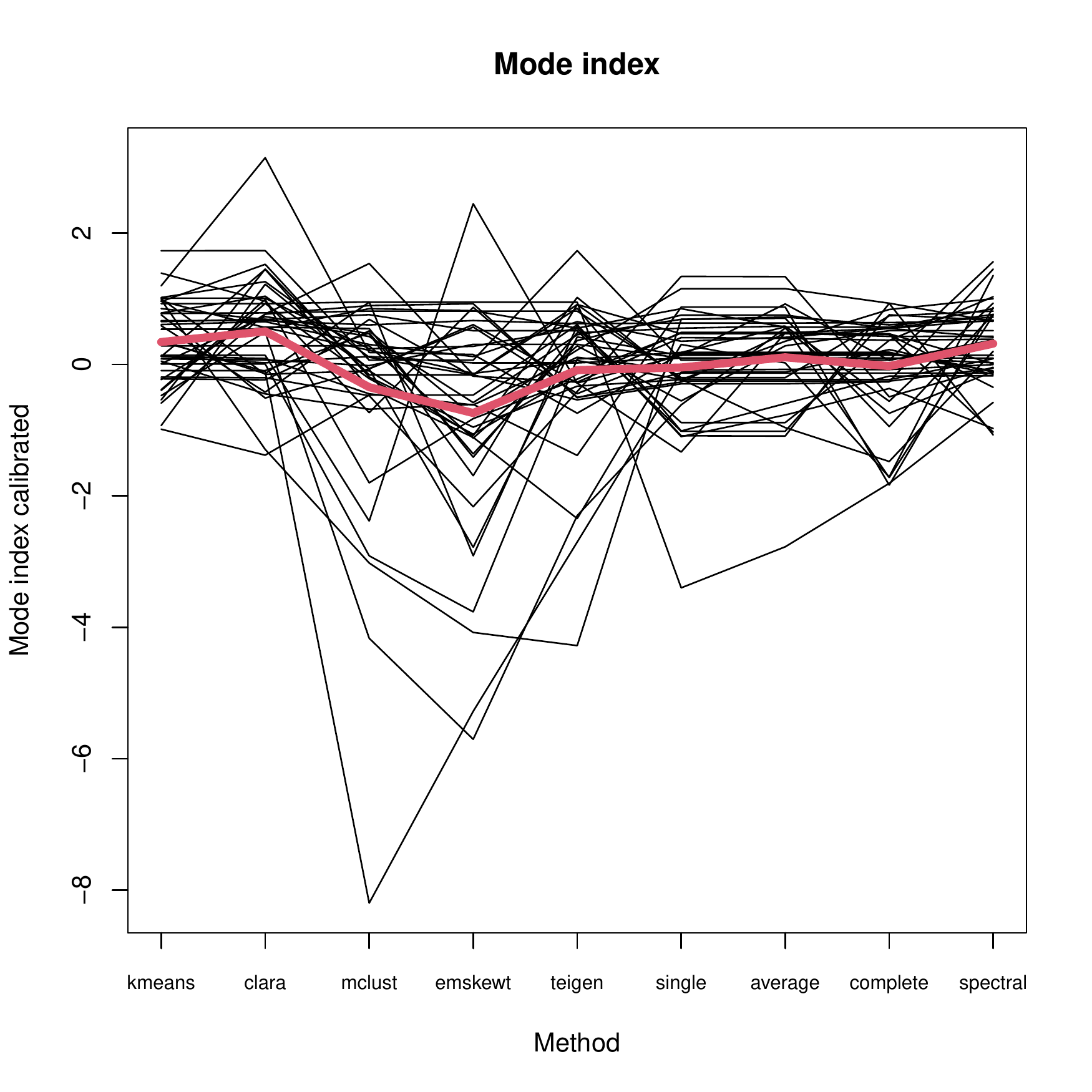}
\end{center}
\caption{Calibrated values of $I_{Pearson\Gamma}^*$ and $I_{dmode}^*$. Values belonging to the same data set are connected by lines. The thick red line gives the
average values.}
\label{fpgdm}       
\end{figure}

\item[Pearson-$\Gamma$] (left side of Figure \ref{fpgdm}): The average results 
for
the methods regarding the representation of the distance structure by the 
clustering vary relatively little compared to the variation over data sets. 
Average linkage
is overall best, and the skew $t$-mixture worst, even if the latter has good
results in some data sets. Single linkage does occasionally very well, but also
worse than the others for a number of data sets.
\item[Density mode index]  (right side of Figure \ref{fpgdm}): Results here
are dominated by variation between data sets as well. 
Interestingly, the methods based on
mixtures of unimodal distributions do not do best here, but rather clara and 
spectral clustering. Once more the mixture of skew $t$-distributions does worst,
with outliers in both directions.

\begin{figure}
\begin{center}
  \includegraphics[width=0.48\textwidth]{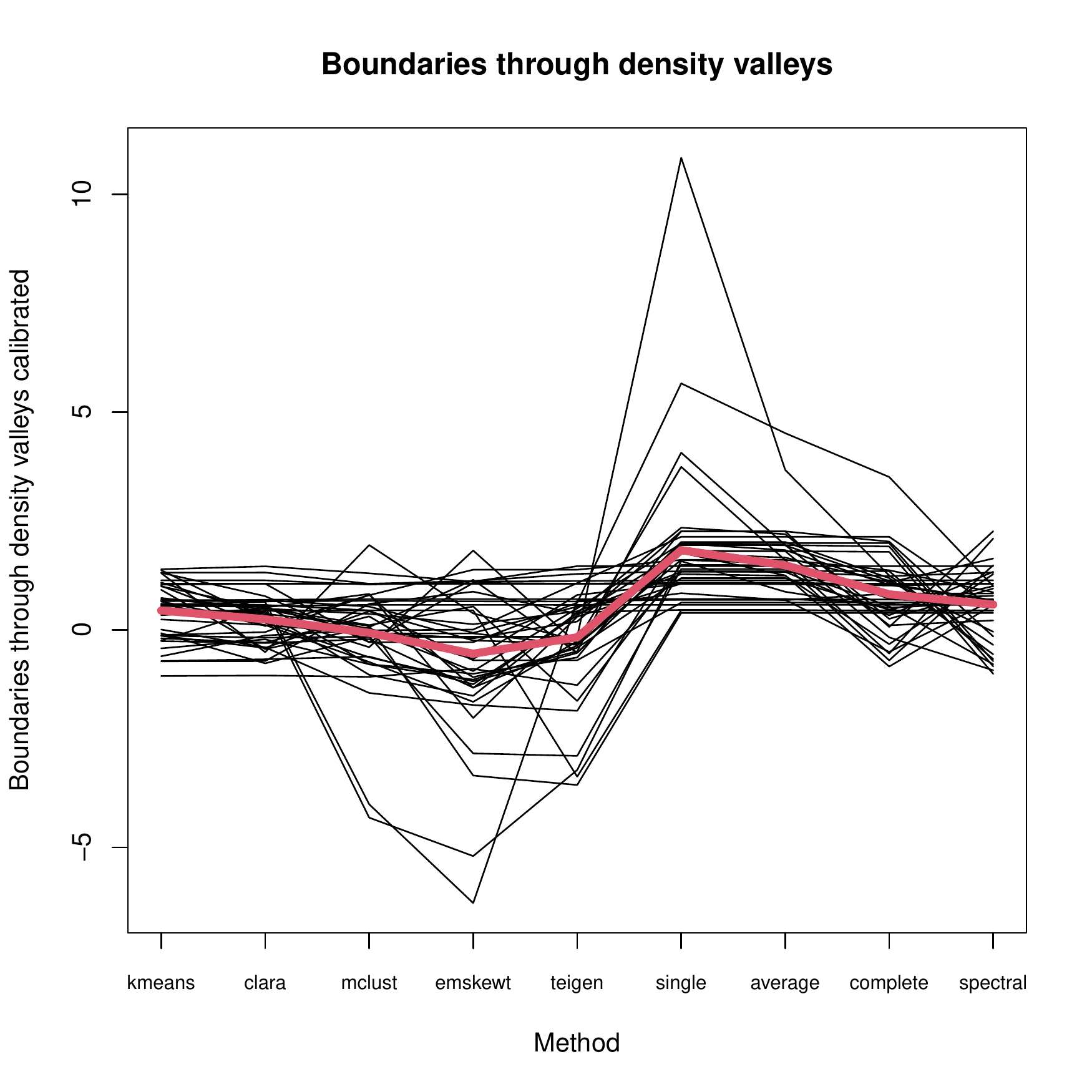}
  \includegraphics[width=0.48\textwidth]{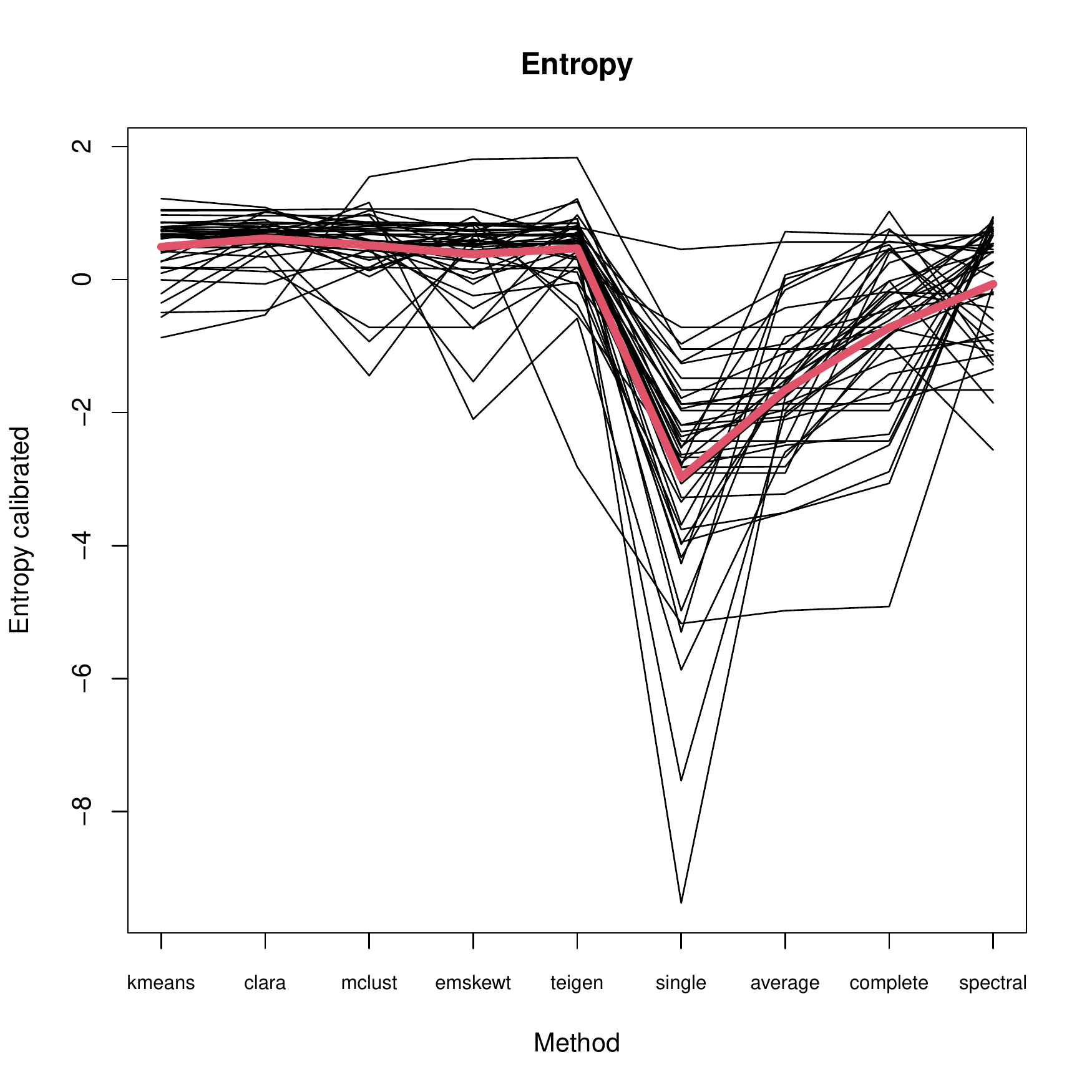}
\end{center}
\caption{Calibrated values of $I_{denscut}^*$ and $I_{entropy}^*$. Values belonging to the same data set are connected by lines. The thick red line gives the
average values.}
\label{fdcen}       
\end{figure}

\item[Density cutting]  (left side of Figure \ref{fdcen}):
Due to its focus on cluster separation, single linkage is best at avoiding
cutting through density mountains. The skew $t$- and $t$-mixture have the 
strongest tendency to put cluster boundaries in high density areas, but 
differences between methods are not large.
\item[Entropy]  (right side of Figure \ref{fdcen}): clara yields
the highest average entropy followed by $K$-means, but differences between these
and the three 
mixture models do not seem significant. This runs counter to the idea, 
sometimes found in the literature, that $K$-means favours similar cluster sizes 
more than mixtures, or even implicitly assumes them. 
The other four methods have a clear tendency to produce less 
balanced clusters, particularly single linkage, but also average and 
complete linkage, and to some lesser extent spectral clustering.

\begin{figure}
\begin{center}
  \includegraphics[width=0.48\textwidth]{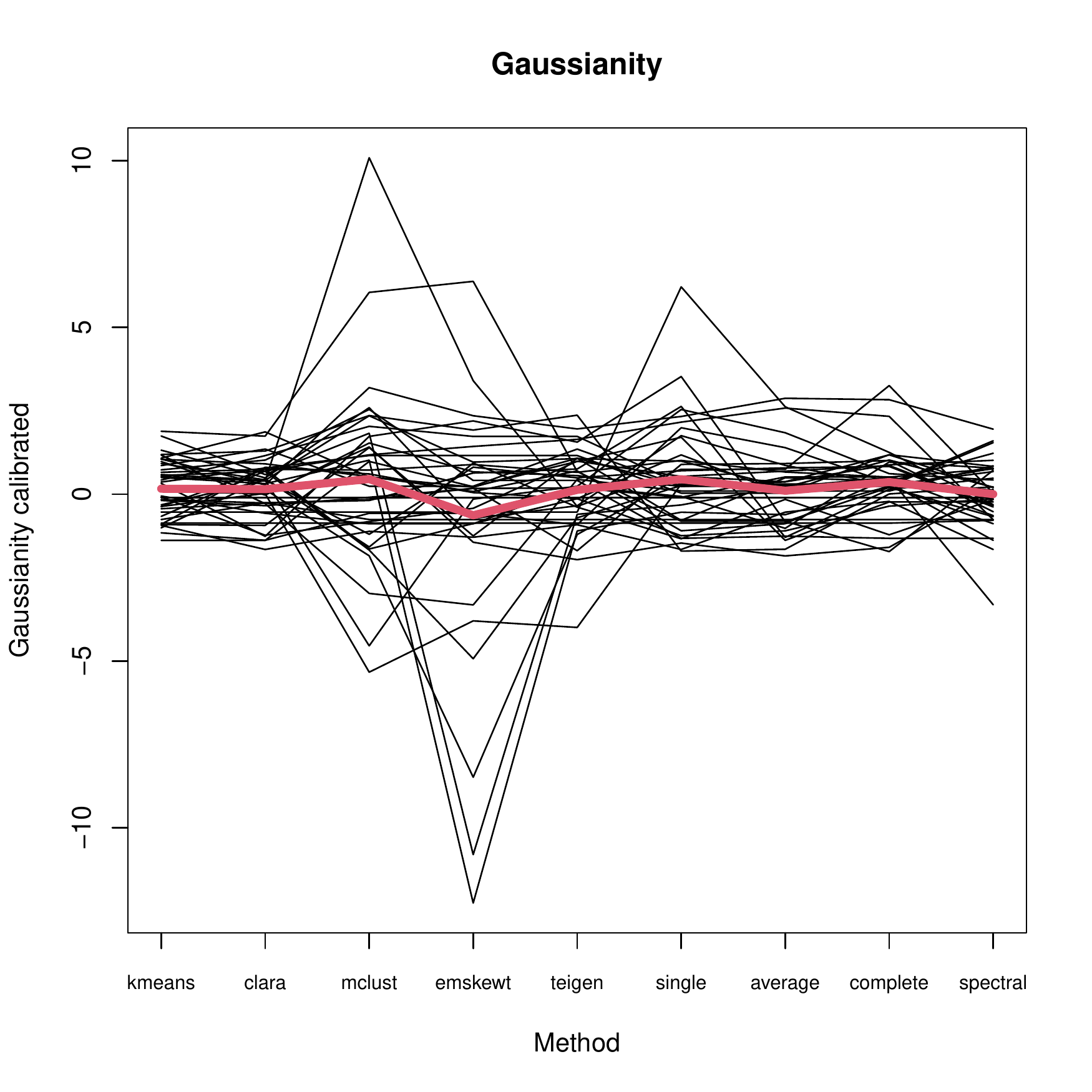}
  \includegraphics[width=0.48\textwidth]{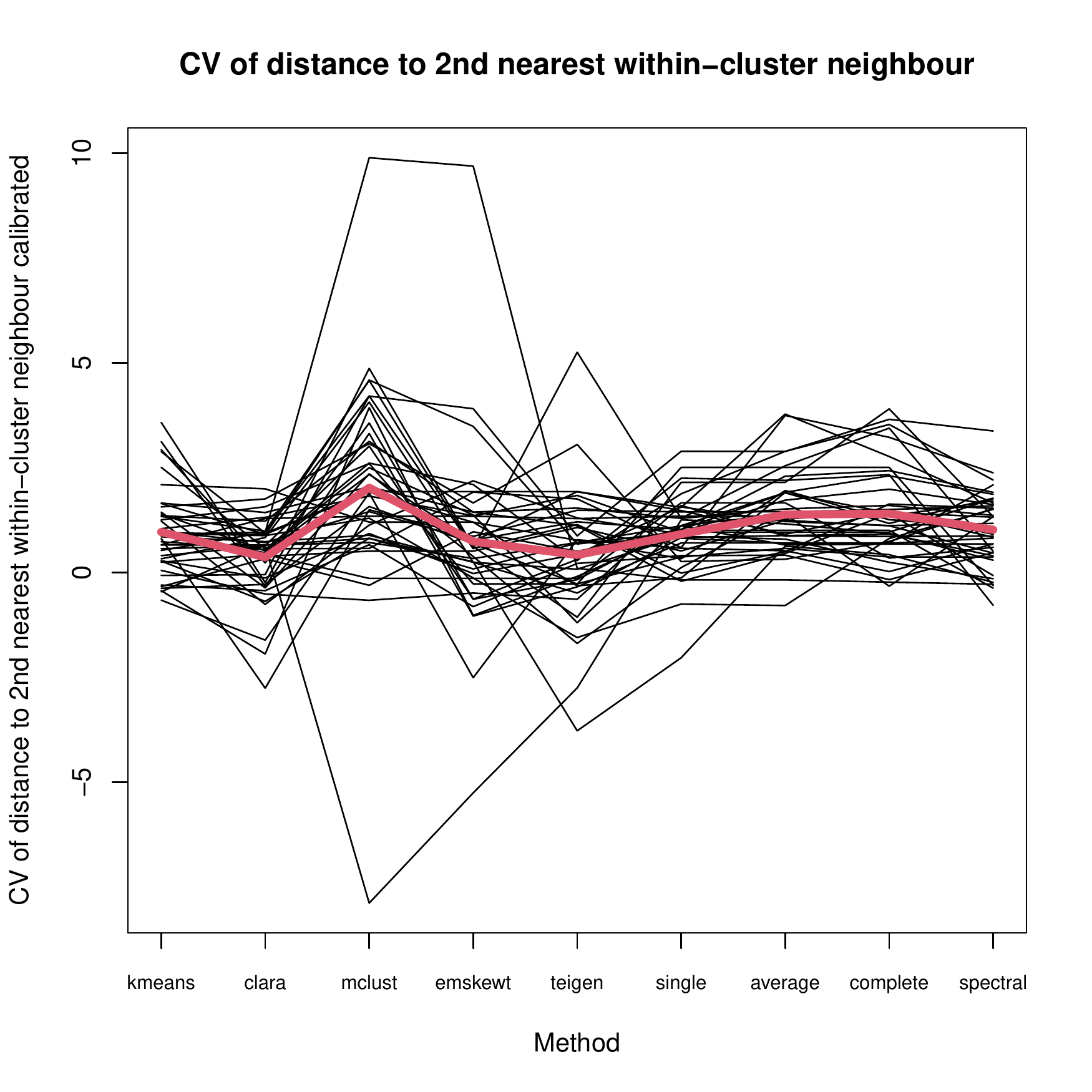}
\end{center}
\caption{Calibrated values of $I_{kdnorm}^*$ and $I_{cvdens}^*$. Values belonging to the same data set are connected by lines. The thick red line gives the
average values.}
\label{fnorcvd}       
\end{figure}

\item[Gaussianity] (left side of Figure \ref{fnorcvd}): Although the Gaussian
mixture produces on average the most Gaussian-looking clusters, as was to be 
expected, the differences between all nine methods look largely insignificant.
The Gaussian mixture has positive and negative outliers, the skew $t$-mixture 
only negative ones. 
\item[CV of distances to within-cluster neighbours]  
(right side of Figure \ref{fnorcvd}): Despite one lower outlier, the Gaussian
mixture tends to produce the largest cvnnd, i.e., the 
lowest within-cluster CVs. 
It probably helps that large variance clusters can bring together 
observations that
have large distances between each other and to the rest. clara and the 
$t$-mixture produce the lowest cvnnd values. Differences
between the other methods are rather small.

\begin{figure}
\begin{center}
  \includegraphics[width=0.48\textwidth]{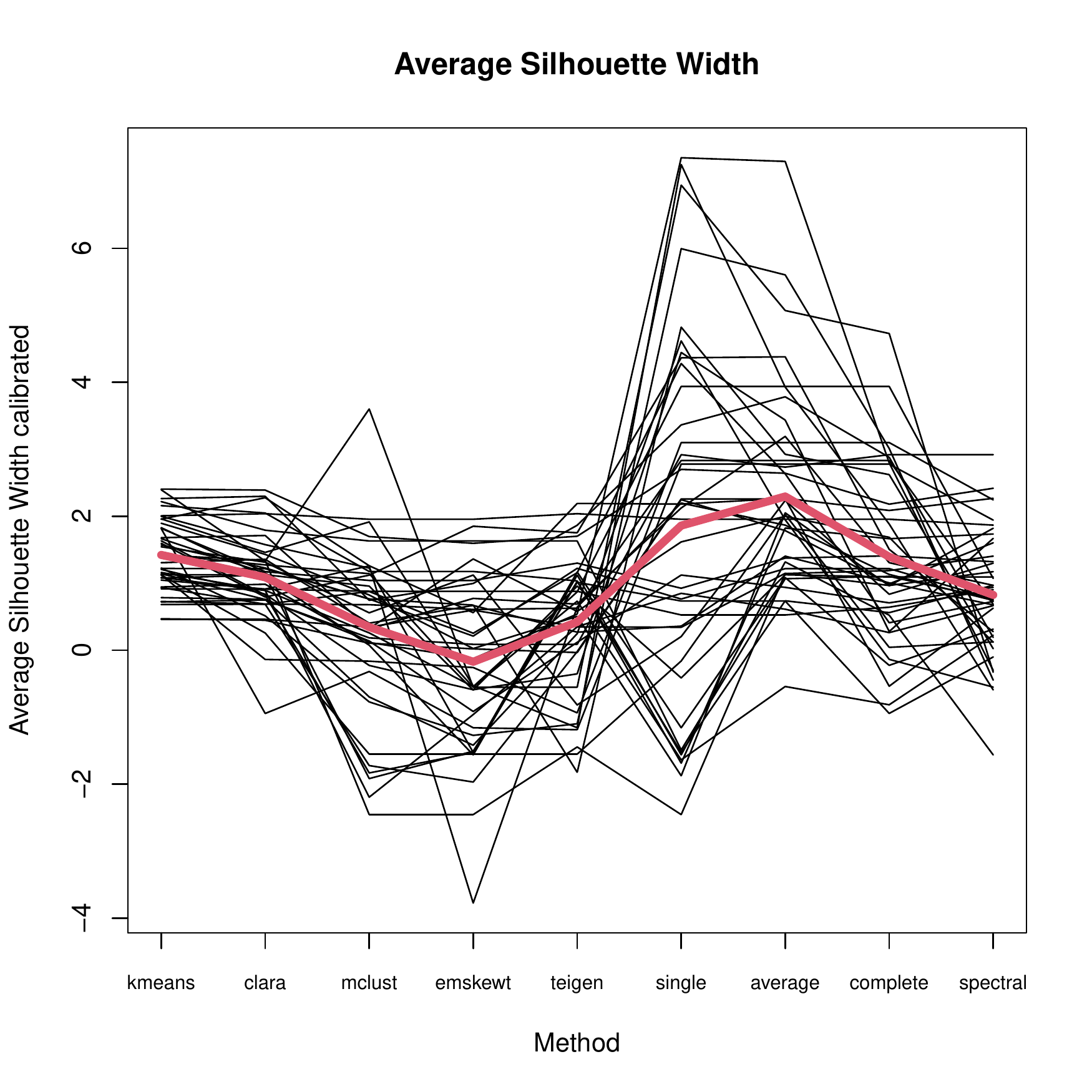}
\end{center}
\caption{Calibrated values of the ASW. Values belonging to the same data set are connected by lines. The thick red line gives the
average values.}
\label{fasw}       
\end{figure}

\item[Average silhouette width]  (left side of Figure \ref{fasw}): Average
linkage is a method that explicitly balances separation and
homogeneity, and consequently it achieves the best ASW values.
$K$-means achieves higher values than
complete linkage, but the remaining methods do worse than the linkage methods.
ASW had been originally proposed for use with clara (\cite{KauRou90}), but
clara does not produce particularly high ASW values, if better than the mixture
models and spectral clustering. 
\end{description}
These results characterise the clustering methods as follows:
\begin{description}
\item[kmeans] clearly favours within-cluster homogeneity over 
separation. It does not favour entropy as strongly as some literature suggests;
in this respect it is in line with clara and the mixture
models, ahead of the remaining methods. It should be noted that entropy
is treated here as a potentially beneficial feature of a clustering, whereas
some literature makes it seem like a defect of kmeans that such solutions 
are favoured (as far as this in fact happens). 
\item[clara] has largely similar characteristics to kmeans. 
It is slightly 
worse regarding the representation of the distance structure and the ASW. 
It is slightly better regarding clusters with density decrease from the 
mode. This may have to do with the fact that the density goes down faster from
the mode for the multivariate Laplace distribution (where the log-likelihood 
sums up unsquared distances) than for the Gaussian distribution (which 
corresponds to squared distances).
\item[mclust] produces clusters with the highest Gaussianity, but only by a
rather insignificant distance. It is best regarding uniformity as measured 
by cvnnd. The reason for this is probably its 
ability to build clusters with large within-cluster variation collecting
observations that have large distances to all or most other points, whereas 
other methods either need to isolate such observations in one-point clusters, 
or integrate them in clusters with denser cores. Mixtures of $t$- and skew 
$t$-distributions could in principle also produce large variance clusters, but
the shapes of $t$- and skew $t$-distributions allow to integrate outlying 
observations more easily with denser regions.

mclust often
tolerates large within-cluster distances, whereas its 
clusters are not on average better separated than those from $K$-means. On the
other hand, its cluster sizes are not significantly less well balanced. Its 
ability to produce clusters with strongly different within-cluster variance 
makes it less suitable regarding Pearson-$\Gamma$ and the ASW,
which treat distances in the same way in all clusters.
\item[emskewt] looks bad on almost all internal indexes. It is not particularly
bad regarding recovery of the ``true'' clusters though, see Section 
\ref{srecovery}. This means that the 
current collection of internal indexes does not capture favourable 
characteristics of 
skewly distributed clusters appropriately; it also means that emskewt is not
an appropriate method for finding clusters with the characteristics that are
formalised by the internal indexes. 
\item[teigen] has a profile that is by and large very similar to the one of 
mclust, apart from being slightly better regarding the maximum diameter, and 
slightly worse regarding Gaussianity and uniformity. 
\item[single linkage] has a very distinct profile. It is best regarding separation, avoiding wide within-cluster gaps, and cluster boundaries through density valleys, and worst by some distance regarding within-cluster homogeneity and entropy.
\item[average linkage] has similar strengths and weaknesses as single linkage,
but not as extreme. It is the best method regarding Pearson-$\Gamma$ and the ASW, both of which balance homogeneity and separation and measure therefore how much the clustering is in line with the distance structure.
\item[complete linkage] is best regarding the maximum diameter. In most other
respects it stands between single and average linkage on one side and the centroid- and mixture-based methods on the other side. 
\item[spectral] is another method that provides a compromise between the rather separation-oriented single and average linkage on one side and the rather homogeneity-oriented centroid- and mixture-based methods. Its maximum cluster diameter is rather high on average. Its mode index value is good if not clearly different from the one of clara. Its mid-range entropy value may look attractive in applications in which a considerable degree of imbalance in the cluster sizes may seem realistic but the tendency of the linkage methods to produce one-point clusters should be avoided.
\end{description}
The multivariate characterisation of the clustering methods also allows to map 
them, using a principal components analysis (PCA). Results of this are shown in 
Figure \ref{fpc}. On the left side, PCs are shown using every index
value for every data set as a separate value, i.e., 42*11 variables. The first 
two PCs carry 30.9\% and 16.6\% of the variance, respectively. 
On the right side, the PCA is performed on 11 variables that give average index
values over all data sets. While this reduces information, it allows to show
the indexes as axes in a biplot. The first 
two PCs here carry 50.0\% and 19.7\% of the variance, respectively. After 
rotation, the maps are fairly similar. Using the more detailed 
data set information, spectral seems much closer to kmeans and clara than to 
mclust and teigen, but the apparent similarity to the latter ones using 
average index values is an effect of dimension reduction; involving
information from the third PC (not shown), the similarity structure is more
similar to that of the plot using all 42*11 variables. The biplot on the right
side shows the opposite tendencies of separation on one hand and entropy
and average within distances on the other hand when characterising the methods,
with indexes such as maximum diameter, density mode, Pearson-$\Gamma$, 
and the ASW opening another dimension, rather corresponding to kmeans, average,
and complete linkage. Qualitative conclusions from these maps
agree roughly with those in \cite{JTLB04}, where more clustering algorithms,
but fewer data sets, were involved.

\begin{figure}
\begin{center}
  \includegraphics[width=0.48\textwidth]{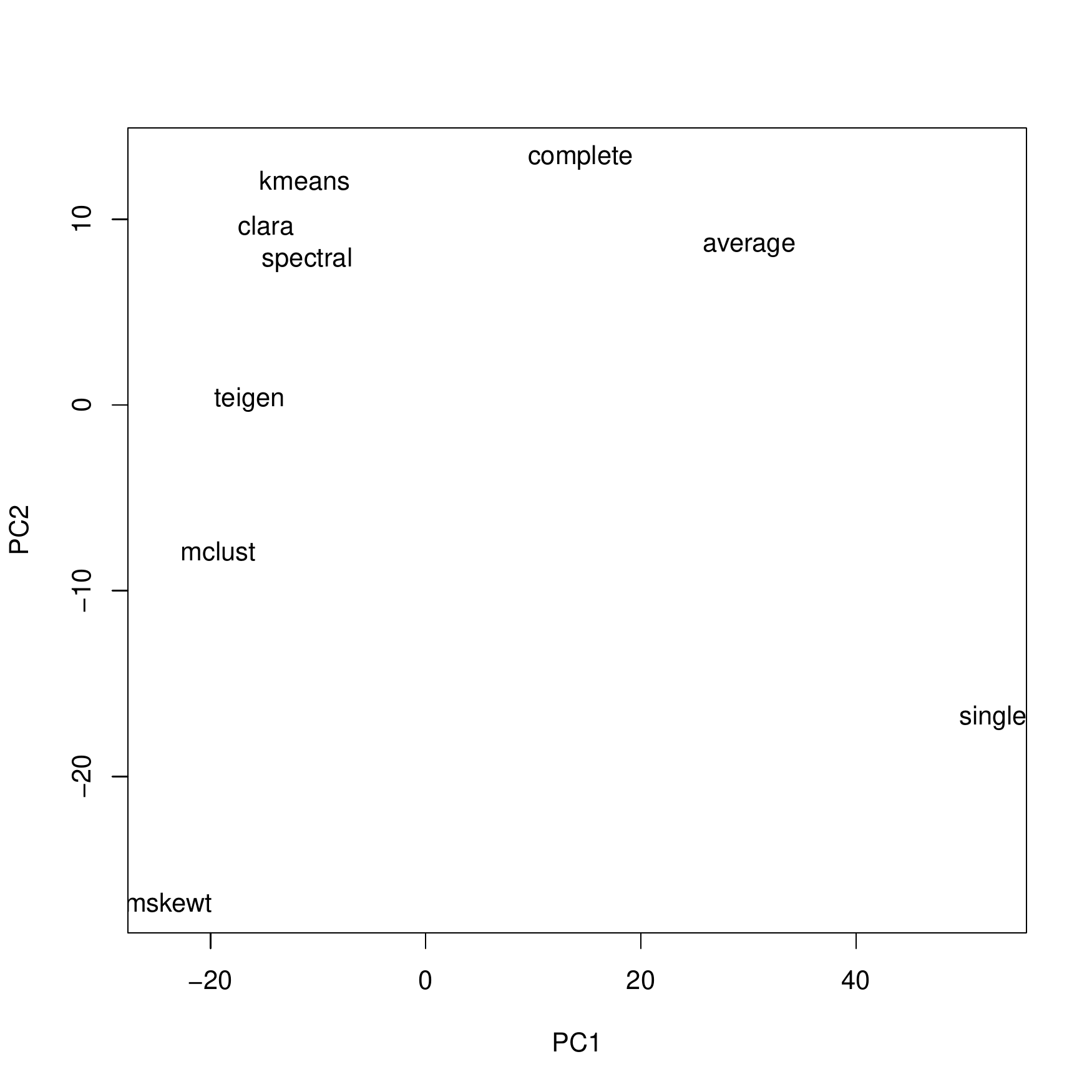}
  \includegraphics[width=0.48\textwidth]{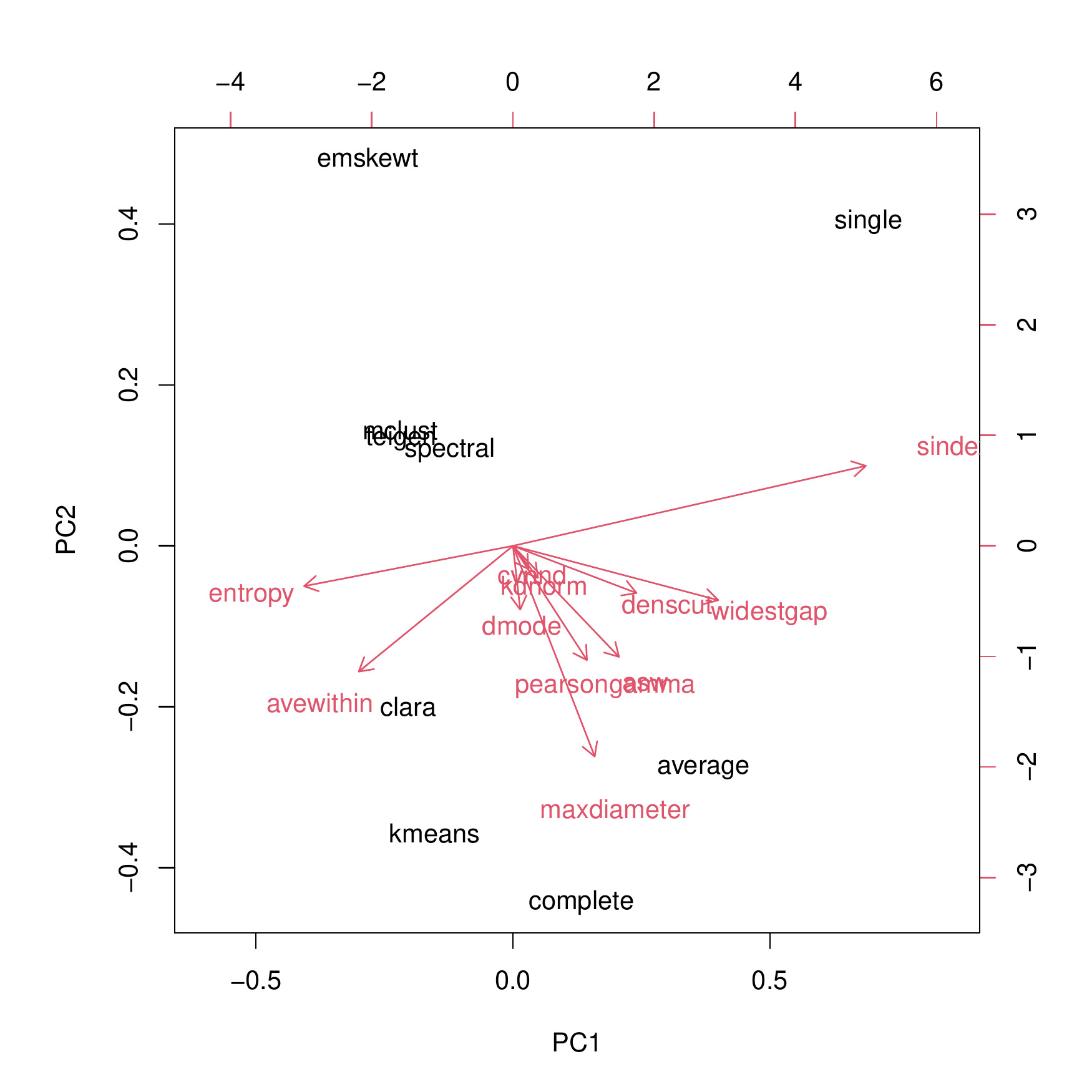}
\end{center}
\caption{Clustering methods mapped on first two principal components from
using all data sets separately (left side), and from using mean values over the data sets (right side).}
\label{fpc}       
\end{figure}

\begin{figure}
\begin{center}
  \includegraphics[width=0.48\textwidth]{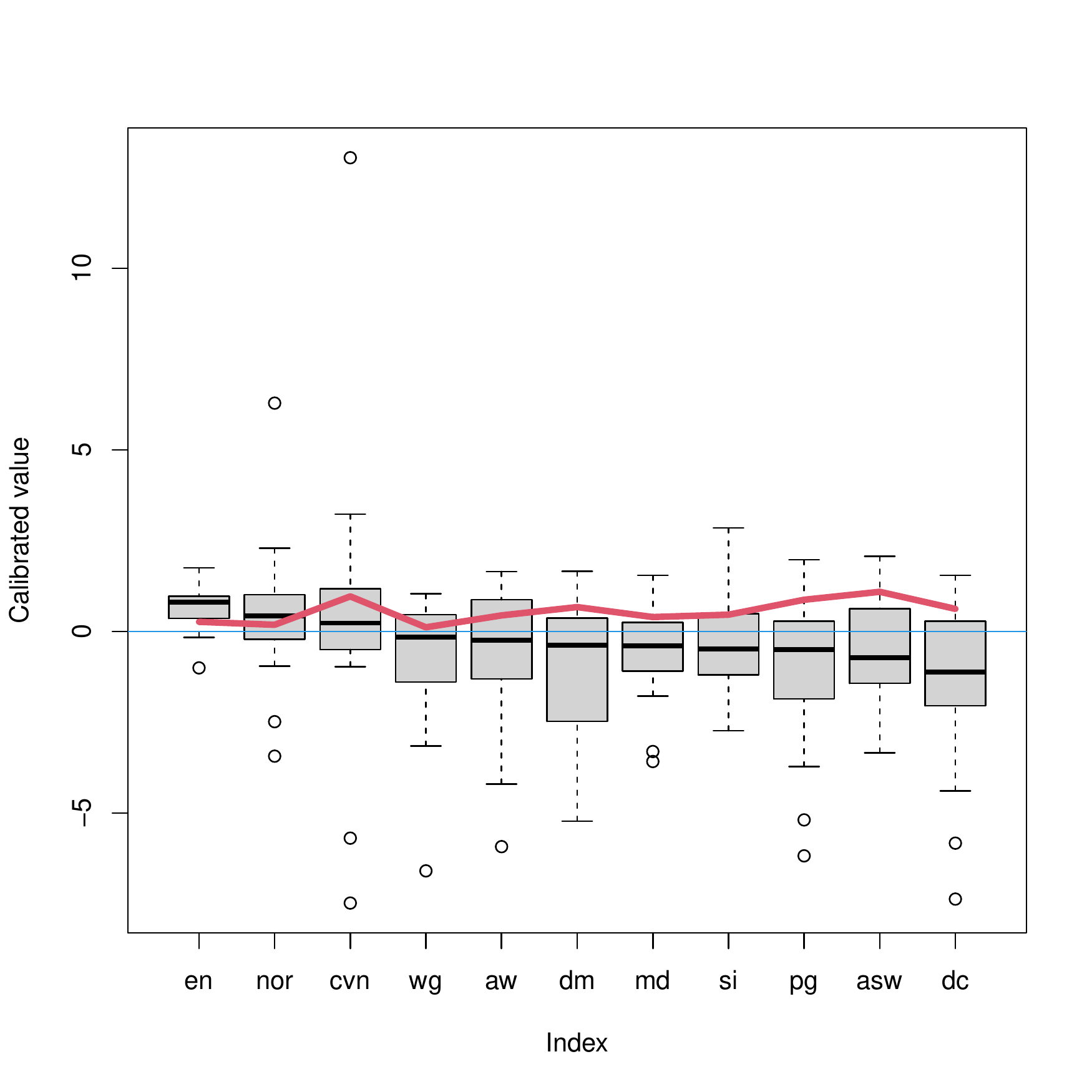}
\end{center}
\caption{The boxplots show the distributions of the internal indexes 
computed on the ``true'' clusterings. The red line shows the average 
index values produced by the clustering methods.}
\label{ftrueindex}       
\end{figure}

The study data allow to also investigate the values of the internal indexes
computed for the ``true'' clusterings. These are shown in Figure 
\ref{ftrueindex}. Only the entropy and Gaussianity are clearly above the 
mean zero of the random clustering ensemble (which includes the solutions from
the proper clustering methods as a small minority), and also above the mean for
the clustering methods. The clustering methods are on average all above zero, 
which should be expected, because these are meant to be desirable features 
of a good clustering, and as such should be better for the proper 
clustering methods than for the random ones. The methods achieve the highest
average for the ASW, 
which makes sense as this attempts to measure general clustering quality.
The fact that index values are mostly
below zero for the ``true'' clusterings can be interpreted in such a way that
many given ``true'' clusterings are data analytically wanting. The high values
for entropy are probably artificial, due to a biased choice of data sets. The 
high values for Gaussianity, however, could suggest that there is a tendency in
some real clusters, i.e., homogeneous subpopulations, to approximate the 
Gaussian distribution. A possible explanation is that in a crisp clustering of 
a data set produced by a clustering method, 
tails of a within-cluster distribution tend to be cut off in the 
direction of other clusters, whereas ``true'' clusters tend to have some 
proper overlap (clearly separated clusters are in my experience indeed rare in
real data), which is in line with the low values of the separation and denscut 
(cluster boundaries running through density valleys) index. This probably also
affects the ASW and Pearson-$\Gamma$.

\subsection{Recovery of ``true'' clusterings}
\label{srecovery}
\begin{figure}
\begin{center}
  \includegraphics[width=0.48\textwidth]{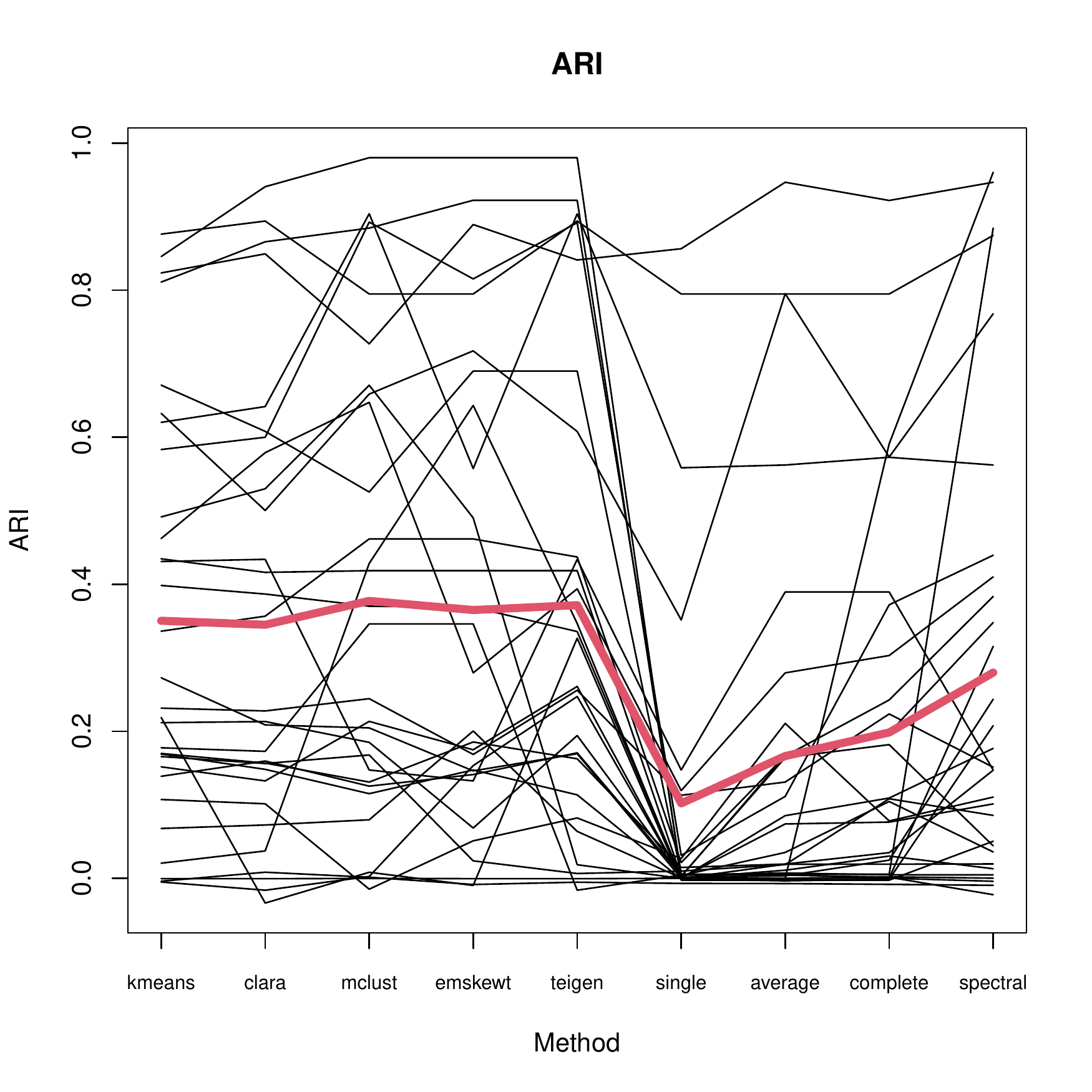}
\end{center}
\caption{Adjusted Rand Index values by method. Values belonging to the same data set are connected by lines. The thick red line gives the
average values.}
\label{fari}       
\end{figure}

The quality of the recovery of the ``true'' clusterings is measured by the ARI, BCubed, and the VI.  Figure \ref{fari} shows the ARI-values achieved by the different clustering methods. On average, there is a clear advantage of the centroid- and mixture-based methods compared with the linkage methods (single linkage is clearly the worst), and spectral clustering is in between. Every method achieves good results on some data sets, but the linkage methods produce an ARI around zero on many data 
sets. Differences between kmeans, clara, mclust, emskewt, and teigen do not seem
significant but are clearly dominated by variation. On some data sets all methods produce very low values, and no method achieves an ARI larger than 0.5 on more than half of the data sets. The mean ARI is 0.28, the mean ARI of the best clusterings for every data set is 0.46. Interpreting these numbers, it has to be kept in mind that the given ``true'' clustering does not necessarily qualify as the best clustering of the data from a data analytic point of view; some of these are neither homogeneous nor separated. Furthermore there may be meaningful clusters in the data that differ from those declared as ``truth''. A better recovery does not necessarily mean that a method delivers the most useful clustering that can be found. On the other hand, some given ``true'' clusterings correspond to clearly visible patterns in the data, and at least some methods manage to find them. Overall, the variation is quite high.

\begin{figure}
\begin{center}
  \includegraphics[width=0.48\textwidth]{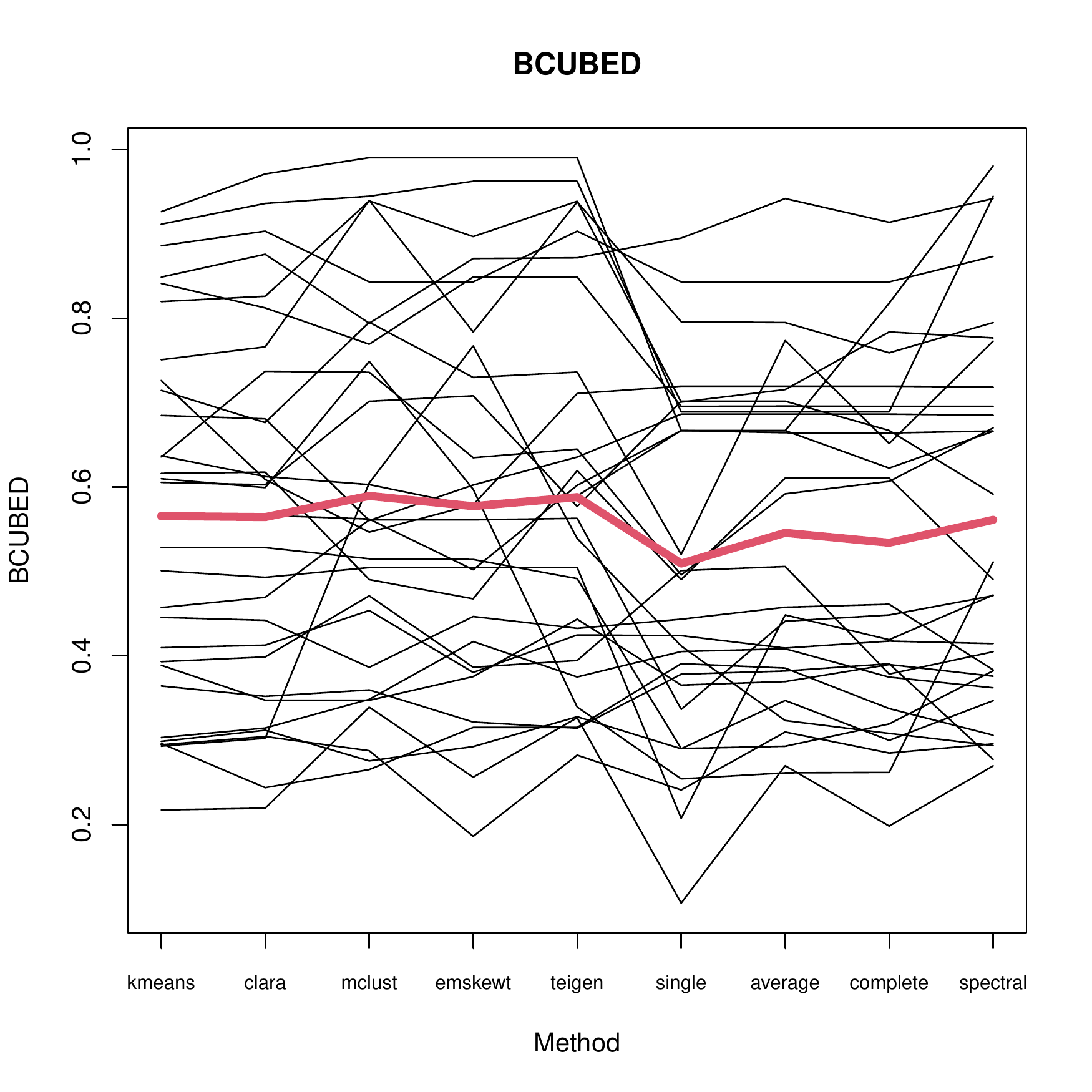}
  \includegraphics[width=0.48\textwidth]{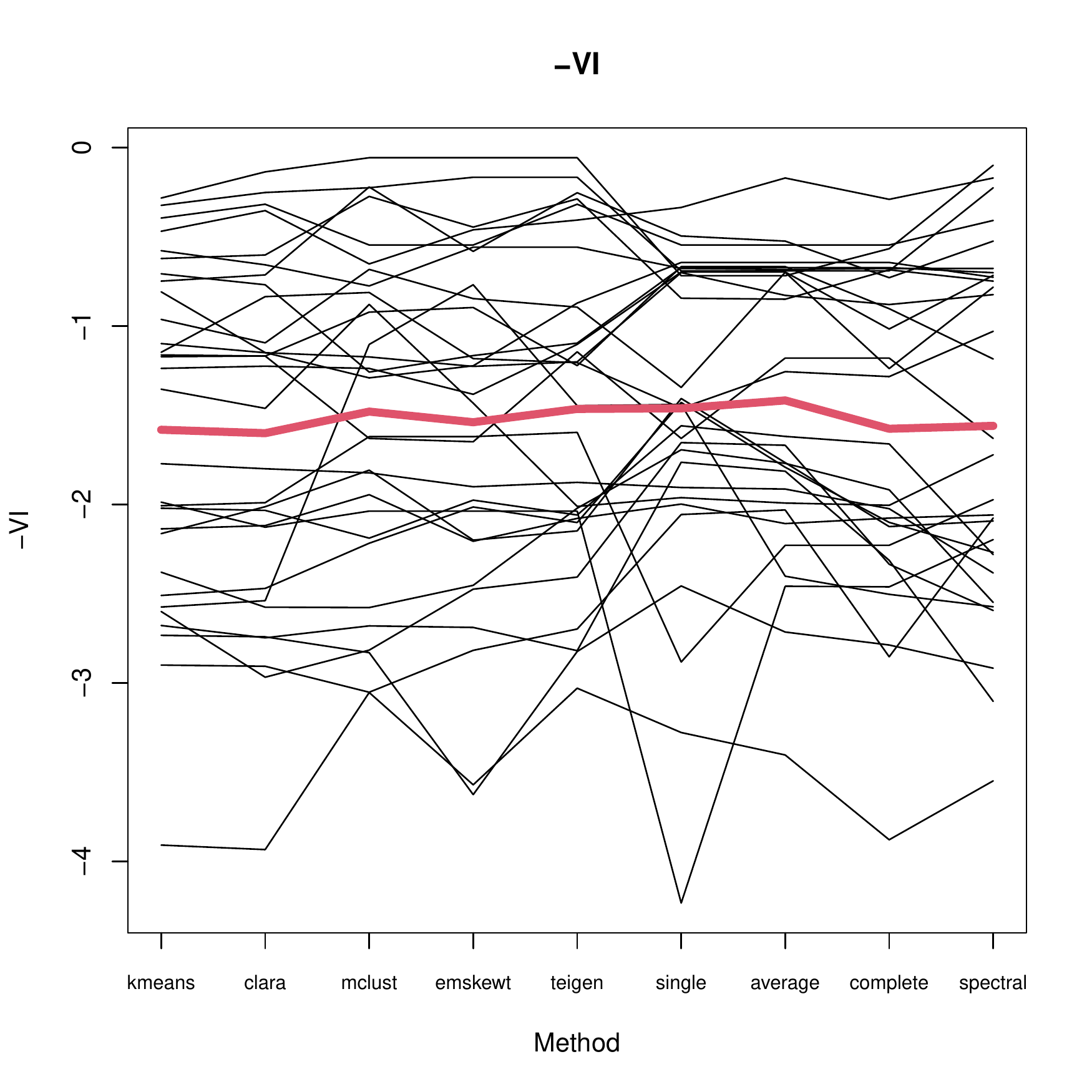}
\end{center}
\caption{BCubed and negative Variation of Information values by method. Values belonging to the same data set are connected by lines. The thick red line gives the
average values.}
\label{fbcvi}       
\end{figure}

The picture changes strongly looking at the results regarding BCubed and particularly VI, see Figure \ref{fbcvi}. BCubed still shows single linkage as the weakest method, but otherwise differences look hardly significant, and according to the VI, the average quality of the methods is almost uniform.   

Further exploratory analysis (not shown) reveals that better values of the
external indexes are systematically associated with lower data dimension $p$
and lower sample size $n$, the latter probably because of confounding with the
correlated dimension. There was no clear interaction with the methods, and
no clear pattern regarding the number of 
clusters $k$.

Table \ref{tbest} shows how often the different methods come out as the best according to the indexes. This portrays mclust as very successful at recovering the ``truth''. Spectral clustering is hardly ever on top, but it has values very close to the best for a number of data sets. Given that emskewt looks so bad regarding the internal indexes in Section \ref{scharacterisation}, its performance regarding the external indexes looks surprisingly good. The most striking difference between the indexes is that single linkage is not the best method for a single data set with respect to the  the ARI, but it is the best for 11 data sets with respect to the VI. This is explored in the following. 

\begin{table}
\begin{tabular}{|r|lllllllll|}
\hline
& \multicolumn{9}{|c|}{Clustering methods}\\
\hline
Index & \tiny{kmeans} & \tiny{clara} & \tiny{mclust} & \tiny{mskewt} & \tiny{teigen} & \tiny{single} & \tiny{average} & \tiny{complete} & \tiny{spectral}\\
\hline
ARI & 3 & 4 & 8 & 5 & 5 & 0 & 3 & 1 & 1\\
BCubed & 2 & 2 & 7 & 5 & 3 & 4 & 4 & 2 & 1\\
VI & 2 & 1 & 6 & 3 & 3 & 11 & 2 & 1 & 1\\
\hline  
\end{tabular}
\caption{Number of times that a method comes out best according to the three external indexes.}
\label{tbest}
\end{table}

\begin{figure}
\begin{center}
  \includegraphics[width=0.65\textwidth]{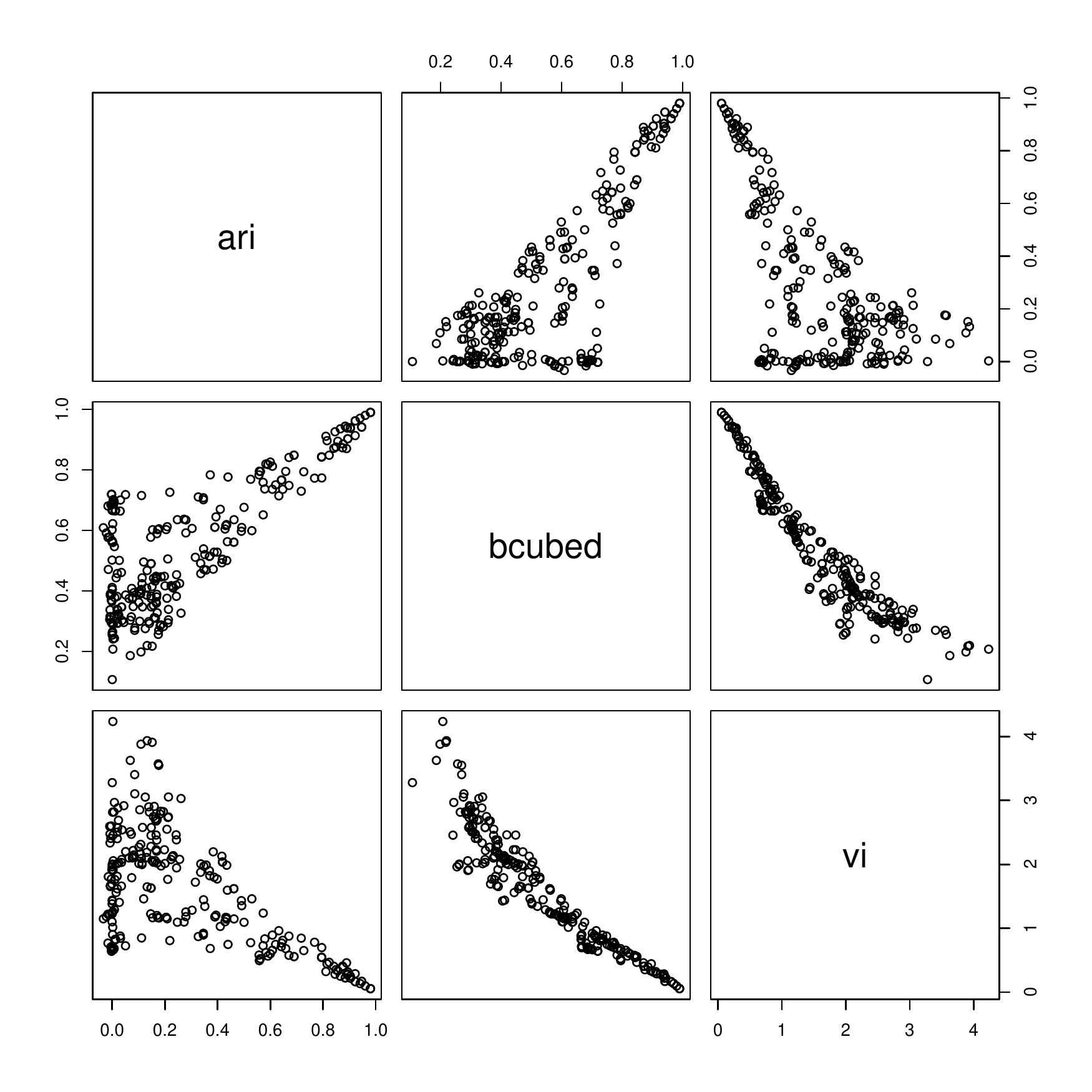}
\end{center}
\caption{Pairs plot of ARI, BCubed, and VI}
\label{fexternal}       
\end{figure}

Figure \ref{fexternal} shows how the three 
indexes are related to each other over all nine clustering methods applied
to the 30 data sets with ``true'' clusterings. VI and BCubed have a 
correlation $\rho$ of -0.94, but the ARI is correlated substantially weaker to 
both,
$\rho=0.75$ with BCubed and $\rho=-0.57$ with VI. 
BCubed can therefore be seen as a compromise 
between the two. In order to explore what causes the differences between 
ARI and VI, in 
Figure \ref{fexternal} it can be seen that the major issue is that the VI
can produce fairly good values close to zero for some situations in which the 
ARI is around zero, indicating unrelated clusterings, or only slightly better.
Generally these situations tend to occur where one clustering is very 
imbalanced, mostly with one or more one-point clusters, whereas
the other one (more often the ``true'' one) is not. The VI involves 
cluster-wise percentages of points 
occurring together in the same cluster in the other clustering, 
and therefore assesses one-point clusters favourably, whereas the random 
labels model
behind ARI indicates that what happens with the object in a one-point
cluster in another (potentially ``true'') clustering is random and therefore 
not meaningful as long as it appears in a substantially bigger cluster there.

For example, consider the data set ``22 - Wholesale'' (see Appendix A2). 
According to the VI, the single linkage clustering is optimal (VI$=0.64$), 
but this has an ARI-value of about 0. It is second best according to BCubed
with a value of 0.72.
Table \ref{tslvi} shows how this 
is related to the ``true'' clustering. 
In favour of this clustering it can be 
said that single linkage cluster 2 is ``pure'' regarding the truth; however, 
it
is clear that any random clustering that fixes one cluster size as 1 will be 
about equally good. This is a rather extreme case, however most of the
assessment differences between ARI and VI (and to a lesser extent  BCubed) 
are of a similar kind. This makes the ARI look like the more appropriate 
index here. 

\begin{table}[tbp]
\caption{Contingency table of ``true'' clustering and single linkage clustering
for data set ``22 - Wholesale''}
\begin{tabular}{|r|ll|}
\hline
& \multicolumn{2}{|c|}{Single linkage cluster} \\
\hline
Truth & 1 & 2 \\
\hline
1 & 297 & 1\\
2 & 142 & 0\\
\hline
\end{tabular}
\label{tslvi}
\end{table}

\subsection{Relating ``true'' cluster recovery to the internal indexes}

\begin{figure}
\begin{center}
  \includegraphics[width=0.8\textwidth]{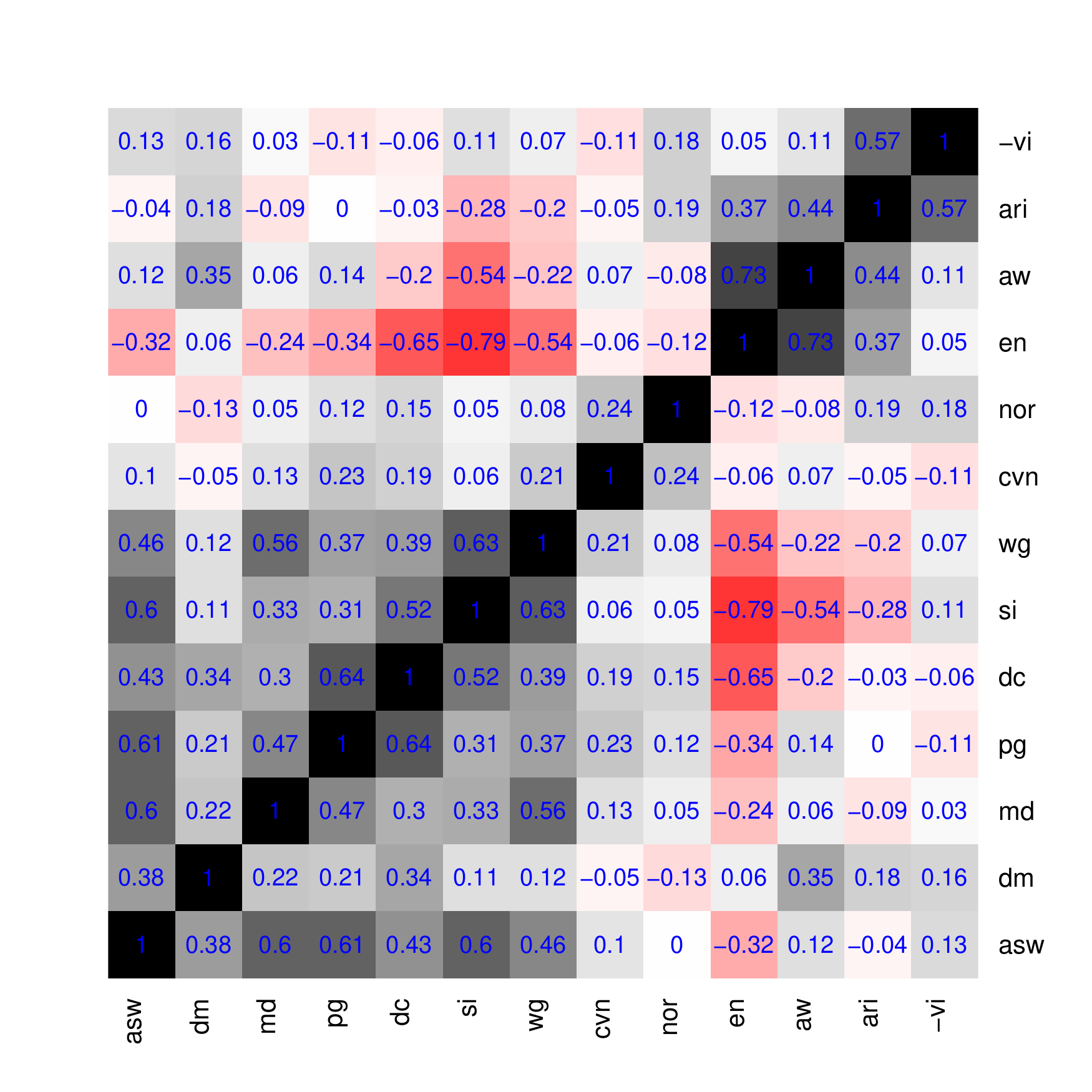}
\end{center}
\caption{Correlation matrix of internal and external validation indexes}
\label{fcor13}       
\end{figure}

It is of interest whether the internal index values, which are observable in a 
real situation, can explain to some extent the performance 
regarding the ``true'' cluster recovery. A tool to assess this is a linear
regression
with an external index as response, and the internal indexes as explanatory
variables. There is dependence between the different clusterings on the same
data set, and this can be appropriately handled using a random data set 
effect. 

An important issue is that the internal indexes are correlated, which can make
the interpretation of the regression results difficult. Figure \ref{fcor13} 
shows the correlation structure among the internal indexes, ARI and -VI (BCubed
is not taken into account in this section due to the high correlation with VI). 
The order of indexes in Figure  \ref{fcor13} was determined by a hierarchical 
clustering using correlation dissimilarity, however -VI and ARI were put on top
due to their different role in the regression, and the ASW was put at the 
bottom. The ASW is not involved in the regression, as it is defined in order to
compromise between homogeneity and separation, which themselves are represented
by other internal indexes. It is involved in Figure  \ref{fcor13} because its 
correlation to the other indexes may be of interest anyway. One thing that 
can be seen is that it is fairly strongly correlated to a number of other 
indexes, particularly maximum diameter, Pearson-$\Gamma$, and the separation 
index, but rather weakly to the average within-cluster distances meant to
formalise homogeneity. 

Considerable correlation occurs between the average within-cluster distances 
and the entropy. Both of these are the internal indexes with the highest 
correlation to the ARI. This is a problem for interpretation because this
means that entropy and homogeneity are confounded when explaining recovery
success. Furthermore, both, entropy in particular, are strongly negatively
correlated with separation, which may explain the negative correlation between
separation and the ARI. There is no further high ($>0.2$) 
correlation between either -VI or ARI and other internal indexes. It is obvious 
that the ARI is closer connected to entropy and homogeneity, whereas the -VI
is more positively connected to separation. There are a number of further 
correlations among the internal indexes; separation, the density mode and
cut indexes, Pearson-$\Gamma$, the maximum cluster diameter, and the absence of
large within-cluster gaps are all positively connected. The Gaussianity index
and the nearest neighbours CV are correlated 0.24 to each other; all their other correlations are lower.

\begin{table}
\caption{Mixed-effects regression results regressing ARI, -VI, respectively,
on the internal indexes excluding the ASW.}
\begin{tabular}{|l|rrr|rrr|}
\hline
Response & \multicolumn{3}{|c|}{ARI} & \multicolumn{3}{|c|}{-VI}\\
\hline
Indexes & Coefficient & $t$ & $p$ & Coefficient & $t$ & $p$ \\
\hline
Intercept & .324 & 6.91 & .000 & -1.54 & -10.11 & .000\\
avewithin & -.019 & -1.34 & .181 & 0.03 & 0.88 & .377\\
maxdiameter & -.025 & -4.03 & .000 &  0.01 & 0.64 & .520\\
widestgap & .014 & 2.00 & .047 & -0.00 & -0.21 & .814\\
sindex & -.010 & -1.65 & .101 & 0.05 & 3.84 & .000\\
pearsongamma &  .020 & 2.43 & .016 & -0.04 & -1.86 & .064\\
dmode & .009 & 0.89 & .374 & 0.05 & 1.92 & .056\\
denscut & .000 & 0.03 & .978 & -0.05 & -1.80 & .074\\
entropy & .088 & 4.69 & .000 & 0.00 & 0.01 & .990\\
kdnorm & .024 & 3.51 & .001 & 0.02 & 1.44 & .151\\
cvnnd & -.006 & -0.86 & .388 & -0.01 & -0.48 & .633\\
\hline
random eff. (data set) & & &.000 & & & .000\\
\hline
\end{tabular}
\label{treg}
\end{table}

Table \ref{treg} gives the results of two regression analyses, with ARI and 
-VI as responses, with a random data set effect. This has been obtained by
the R-package \verb|lme|, \cite{PinBat00}. $p$-values are interpreted in 
an exploratory manner, as they are not precise. However, the null hypotheses of
zero effect of a variable given all other variables are in the model are
of interest here. 

The ARI regression has maximum diameter, entropy, and Gaussianity as highly
significant effects; Pearson-$\Gamma$ is clearly 
significant at 5\%-level. widestgap is borderline significant, which is 
potentially not meaningful given the number of tests. 

The interpretation of entropy (which has the clearly 
largest $t$-value) is problematic for two reasons. Firstly, due to correlation,
its coefficient may partly carry information due to avewithin. Secondly,
eight data sets have artificially balanced classes, which may favour entropy 
among good clusterings. The regression was re-run excluding those data sets
(not shown), yielding by and large the same significances including entropy, but
its $t$-value fell to 2.75. Even in this scenario it cannot be excluded that
the ``sample'' of data sets with known ``true'' clusters favours entropy 
artificially. Gaussianity seems to be a valuable predictor for recovery of 
``true'' classes. The maximum diameter has a negative coefficient, meaning
that on average and controlled for all other indexes, a larger (therefore worse)
maximum cluster diameter went with a better ``truth'' 
recovery regarding the ARI. It is
however clearly correlated with Pearson-$\Gamma$ and widestgap, which have 
positive effects.

Despite a positive relationship between ARI and -VI, the results of
the VI-regression are very different, mainly because -VI can achieve 
high values for clusterings with very low entropy even if the ``true''
clustering is balanced. This means that there is no bias in favour 
of entropy by the data set sample; rather the VI seems biased against 
entropy by definition, see above. The only
clearly significant index for -VI is the separation index, with a positive
coefficient, which was not significant in the ARI-regression. 

Plots of the fitted values of both regressions against their response variable
(not shown) look satisfactorily linear. In principle, the regressions could be
used to predict the ARI or VI for data sets with unknown ``truth'' from the
observable internal indexes, but this will not work very well, due the strong
data set effect.

Overall these results do not allow clear cut conclusions, due to  
correlation, issues with the representativity of the data sets, 
and the very different patterns observed for ARI and VI.
The character of the ``true'' clusterings may just be so diverse that no
general statement about which clustering characteristics allow for good recovery
can be made. Preferring the ARI as external index, the only safely interpretable
significance seems to be the one of Gaussianity, due to its low correlation
with other indexes. 
Separation seems to help in terms of the VI, but this 
includes favouring clusterings that separate outliers as one-point clusters,
arguably an issue with the VI.
   
\section{Discussion}
\label{sdiscussion}
The aim of this study is to characterise the clustering methods in terms of the
internal indexes, to learn about the recovery of ``true'' clusterings, both 
regarding the methods, and regarding characteristics that could be connected
to recovery. 

Regarding the characterisation of the clustering methods, the right side of
Figure \ref{fpc} is probably most expressive, locating the clustering methods 
relative to the internal indexes. Some indexes do not separate the methods
very strongly. Single linkage stands out as being quite different to most other
methods in many respects. On the other hand, the centroid-based methods, the
mixture-based methods and spectral clustering have much in 
common; one surprising result is that $K$-means does not favour balanced 
cluster sizes particularly strongly, compared to the mixture-based methods.
Another result is that single and complete linkage are not opposite extremes,
but rather that on most characteristics of single linkage, complete linkage
is closer to single linkage, with average linkage in between, than the centroid-
based and mixture-based methods. Gaussian mixture-based clustering stands out 
more by its good value regarding uniformity (cvnnd) than
regarding Gaussianity of the clusters. 

Regarding the recovery of ``true'' clusterings, there is large variation 
between the data sets. According to the ARI and BCubed, the Gaussian mixture
is the best for the largest number of data sets. Single Linkage does badly 
regarding the ARI. Differences between the other methods are not that 
pronounced, and all of them did best in some data sets. This includes the skew
$t$-mixture, which does not look good according to the internal indexes but
better regarding the external indexes. There is currently no index, at least
in the collection used here,
that formalises in which sense such a mixture can yield a good clustering. This
is a topic for further work. According to the VI (and to some extent BCubed),
single linkage does much better, but this rather indicates a problem with the
indexes than a good performance of single linkage.

Explaining the ``true'' cluster recovery by the internal indexes does not 
deliver very clear results, except that Gaussianity seems to help, which is 
sometimes achieved by the Gaussian mixture, but only insignificantly more
often than by some other methods. A critical interpretation could be that 
quality according to the internal indexes does not really measure what is 
important for recovery. On the other hand one could argue that this shows
the heterogeneity of ``true'' clusterings, and that there is no ``one
fits it all approach'', neither for clustering, nor for measuring clustering 
quality. The given ``true'' clusterings are of such a nature that their 
recovery cannot be reliably predicted from observable cluster characteristics.

Some problems were exposed with the non-representativity of the data sets,
with ``true'' clusterings, and with the VI (and somewhat less extreme the
BCubed) index. These problems are not exclusive to the present study, and it
can be hoped that these issues are on the radar whenever such benchmark studies
are run. These problems affect analyses involving the ``true
clusterings'' in particular. There is no reason to believe that the results
regarding the internal validation indexes are biased for these reasons.


\section*{Appendix}
\subsection*{A1: Computational details}
The following amendments were made to the clustering functions listed in
Section \ref{sclustering}:
\begin{description}
\item[Mixture models:] Crisp partitions have always been enforced by assigning
observations to the cluster with maximum posterior probability of membership. 
\item[kmeans:] This was run with parameter \verb|runs=10|, governing the number
of random initialisations. The default value is  \verb|runs=1|, which yields 
very unstable results.
\item[emskewt] The function \verb|EmSkew| would occasionally produce errors or 
invalid results. 
It is run inside a wrapper function that enforces a solution in the following
way: For each covariance matrix model\footnote{The shape of a 
skew t-distribution
is defined by the covariance matrix of an involved Gaussian mixture, see 
\cite{LeeMcL13}, although this is not the covariance matrix of the resulting
skew t-distribution.}, starting from (1) the fully flexible model, 
5 attempts (different random initialisations) 
are made to find a solution. If all attempts for a model fail, a
less flexible model is tried out, in the order (2) diagonal covariance matrices,
(3) flexible but equal covariance matrices, 
(4) equal diagonal covariance matrices, (5) equal spherical covariance matrices,
until a valid solution is found. If none of these is successful, the same 
routine is carried out with a mixture of skew normal distributions, and if this
does not yield a valid solution either, \verb|mclustBIC| is called with default 
settings.
\item[teigen] The function \verb|teigen| would occasionally produce errors or
invalid results. 
It is run inside a wrapper function that enforces a solution in the following
way: If no valid solution is found, the wrapper-function for \verb|EmSkew| as
explained above is called, but with \verb|dist="mvt"|, fitting a 
multivariate t-distribution.
\item[specc] The function \verb|specc| would occasionally produce errors or
invalid results. It is run inside a wrapper function. 10 attempts (different
random initialisations) are made to find a solution. If they all fail, all
observations are assigned to cluster 1. While this approach 
may seem unfair for spectral clustering in comparison to \verb|EmSkew|, which
ultimately calls \verb|mclust| and can as such still produce a reasonable 
clustering, the motivation is that a Gaussian mixture model can be seen as a 
constrained version of a mixture of skew t-distributions, whereas spectral 
clustering has no straightforward constrained version that can guarantee a
valid solution. 
\end{description}
In principle there can be situations in which also \verb|mclustBIC| fails to 
deliver a valid solution, however such a situation did not occur in the study.
Exhausting all attempts, both \verb|specc| and \verb|EmSkew| failed twice
before resorting to a one-cluster solution or \verb|mclustBIC|, respectively, 
and \verb|teigen| failed 5 times; in all of these cases \verb|EmSkew| with
\verb|distr="mvt"| delivered a valid solution.

\subsection*{A2: More details on data sets}
Tables \ref{tdata1} and \ref{tdata2}
give a list of the data sets used in the study.

\begin{table}[tbp]
\tiny
\caption{Overview of data sets used in the study. As ``Source'', the source is 
given from which the data set was retrieved for the study, which in some cases 
is not the original source (most data sets retrieved from 
www.openml.org and many from R-packages are from UCI). Missing references:
(i) Turing Institute, Glasgow, (ii) www.bundestag.de (iii) maps.met.police.uk/tables.htm}
\begin{tabular}{|l|l|r|r|r|r|r|r|}
\hline
Number & Name & $n$ & $p$ & $K$ & ``Truth'' given & Source & Reference\\  
\hline
1 & Crabs & 200 & 5 & 4 & Yes & R-MASS & \cite{CamMah74} \\
\hline \multicolumn{8}{|c|}{Morphological measurements of crabs, two species, two sexes}\\ 
\hline
2 & Dortmund & 170 & 5 & 5 & No & See reference & \cite{SomWei05} \\
\hline \multicolumn{8}{|c|}{Various characteristics of the districts of the city of Dortmund}\\ \hline 
3 & Iris & 150 & 4 & 3 & Yes & R-datasets & \cite{Anderson35}\\
\hline \multicolumn{8}{|c|}{Measurements on 50 flowers from each of 3 species of iris}\\ \hline
4 & Vowels & 990 & 10 & 11 & Yes & See reference & \cite{HaTiFr01}\\
\hline \multicolumn{8}{|c|}{Recognition of British English vowels}\\ \hline
5 & Bats & 2677 & 72 & 8 & Yes & V. Zamora-Gutierrez & \cite{Zamora16}\\
\hline \multicolumn{8}{|c|}{Acoustic identification of Mexican bat species}\\ \hline
6 & USArrests & 50 & 4 & 2 & No & R-datasets & \cite{McNeil77}\\
\hline \multicolumn{8}{|c|}{Arrests per 100,000 residents for various crimes in US states 1973}\\ \hline
7 & OliveOil & 572 & 8 & 9 & Yes & R-pdfcluster & \cite{FoArLaTi83}\\
\hline \multicolumn{8}{|c|}{Chemical decomposition of Italian olive oils from 9 regions}\\ \hline
8 & OldFaithful & 299 & 2 & 3 & No & R-MASS & \cite{AzzBow90}\\
\hline \multicolumn{8}{|c|}{Duration and waiting times for eruptions of Old Faithful geyser}\\ \hline
9 & Tetragonula & 236 & 4 & 9 & Yes & R-prabclus & \cite{FCGRO04}\\
\hline \multicolumn{8}{|c|}{Genetic information on 9 species of tetragonula bees}\\ \hline
10 & Thyroid & 215 & 6 & 3 & Yes & R-mclust & \cite{CoBrJoMa83}\\
\hline \multicolumn{8}{|c|}{Results of five laboratory tests diagnosing thyroid gland patients}\\ \hline
11 & Spam & 4601 & 57 & 2 & Yes & R-kernlab & \cite{HaTiFr01}\\
\hline \multicolumn{8}{|c|}{Email spam classification from word and character frequencies}\\ \hline
12 & Wisconsin & 569 & 30 & 2 & Yes & UCI & \cite{StrWolMan93}\\
\hline \multicolumn{8}{|c|}{Diagnosis of breast cancer, measurements of features of image}\\ \hline
13 & Yeast & 1484 & 8 & 10 & Yes & UCI & \cite{HorNak96}\\
\hline \multicolumn{8}{|c|}{Discriminative features for protein Localization Sites in cells}\\ \hline
14 & Vehicle & 846 & 18 & 4 & Yes & R-mlbench & (i)\\
\hline \multicolumn{8}{|c|}{Recognising vehicle type from  silhouettes}\\
\hline
15 & Letters & 2000 & 16 & 26 & Yes & R-mlbench & \cite{FreSla91}\\
\hline \multicolumn{8}{|c|}{Recognising handwritten letters from pixel displays}\\ \hline
16 & Bundestag & 299 & 5 & 5 & No & R-flexclust & (ii)\\
\hline \multicolumn{8}{|c|}{German Bundestag election results 2009 of 5 major parties by constituency}\\ \hline
17 & Finance & 889 & 4 & 2 & Yes & R-Rmixmod & \cite{duJSev10}\\
\hline \multicolumn{8}{|c|}{Predicting firm bankruptcy from four financial ratios}\\ \hline
18 & BankNotes & 200 & 6 & 2 & Yes & R-mclust & \cite{FluRie88}\\
\hline \multicolumn{8}{|c|}{Identifying counterfeit Swiss bank notes from measurements}\\ \hline
19 & StoneFlakes & 79 & 8 & 3 & No & Thomas Weber & \cite{Weber09}\\
\hline \multicolumn{8}{|c|}{Measurements on prehistoric stone tools}\\ \hline
20 & Leaf & 340 & 14 & 30 & Yes & UCI & \cite{SiMaAl12}\\
\hline \multicolumn{8}{|c|}{Shape and consistency measurements on leafs from 30 plant species}\\ \hline
21 & London & 32 & 9 & 4 & No & See reference & (iii) \\
\hline \multicolumn{8}{|c|}{Relative numbers of various crimes in the boroughs of London 2014}\\ \hline
\end{tabular}
\label{tdata1}
\end{table}

\begin{table}[tbp]
\tiny
\caption{Overview of data sets used in the study (part 2). 
As ``Source'', the source is 
given from which the data set was retrieved for the study, which in some cases 
is not the original source (most data sets retrieved from www.openml.org and many from R-packages are from UCI). Missing references: (i) Deepraj Baidya (ii) Dukascopy Historical Data Feed (iii) www.decathlon2000.com}
\begin{tabular}{|l|l|r|r|r|r|r|r|}
\hline
Number & Name & $n$ & $p$ & $K$ & ``Truth'' given & Source & Reference\\  
\hline
22 & Wholesale & 440 & 7 & 2 & Yes & \verb|www.openml.org| & \cite{Abreu11}\\
\hline \multicolumn{8}{|c|}{Spending on various product categories by clients of wholesale distributor}\\ \hline
23 & Heart & 200 & 13 & 5 & Yes & \verb|www.openml.org| & \cite{DJSPSSGLF89}\\
\hline \multicolumn{8}{|c|}{Diagnosing different stages of heart disease by diagnostic measurements}\\ \hline
24 & MachineKnow & 403 & 5 & 5 & No & \verb|www.openml.org| & \cite{KaSaCo13}\\
\hline \multicolumn{8}{|c|}{Students' knowledge status about the subject of Electrical DC Machines}\\ \hline
25 & PlantLeaves & 1599 & 64 & 100 & Yes & \verb|www.openml.org| & \cite{Yanetal13}\\
\hline \multicolumn{8}{|c|}{Plant species classification by texture detected from leaf images}\\ \hline
26 & RNAYan & 90 & 2 & 7 & Yes & Bioconductor & \cite{Yanetal13}\\
\hline \multicolumn{8}{|c|}{RNA sequencing data distinguishing cell types in human embryonic development}\\ \hline
27 & RNAKolo & 704 & 5 & 3 & Yes & Bioconductor & \cite{Kolodziejczyketal15}\\
\hline \multicolumn{8}{|c|}{RNA sequencing data on mouse embryonic stem cell growth}\\ \hline
28 & Cardiotocography & 2126 & 23 & 10 & Yes & \verb|www.openml.org| & \cite{ABGMP00}\\
\hline \multicolumn{8}{|c|}{Classification of cardiotocograms into pattern classes}\\ \hline
29 & Stars & 240 & 4 & 6 & Yes & Kaggle & (i) \\
\hline \multicolumn{8}{|c|}{Predict star types from features of stars}\\ \hline
30 &  Kidney & 203 & 11 & 2 & Yes & R-teigen & \cite{UCI17} \\
\hline \multicolumn{8}{|c|}{Presence or absence of chronic kidney disease from diagnostic features}\\ \hline
31 & BreastTissue & 106 & 9 & 4 & Yes & \verb|www.openml.org| & \cite{Jossinet96}\\
\hline \multicolumn{8}{|c|}{Classes of breast carcinoma diagnosed by impedance measurements}\\ \hline
32 & FOREX & 1832 & 10 & 2 & Yes & \verb|www.openml.org| & (ii) \\
\hline \multicolumn{8}{|c|}{Historical price data EUR/JPY for predicting direction next day}\\ \hline
33 & SteelPlates & 1941 & 24 & 7 & Yes & \verb|www.openml.org| & \cite{Buscema98}\\
\hline \multicolumn{8}{|c|}{Classification of steel plates faults from various measurements}\\ \hline
34 & BostonHouse & 506 & 13 & 5 & No &  \verb|www.openml.org| & \cite{HarRub78}\\
\hline \multicolumn{8}{|c|}{Multivariate characterisation of different areas of Boston}\\ \hline
35 & Ionosphere & 351 & 32 & 2 & Yes & \verb|www.openml.org| & \cite{SWHB89}\\
\hline \multicolumn{8}{|c|}{Radar data to distinguish free electron patterns from noise in ionosphere}\\ \hline
36 & Glass & 214 & 9 & 6 & Yes & R-MASS & \cite{VenRip02}\\
\hline \multicolumn{8}{|c|}{Identify type of glass from chemical analysis}\\ \hline
37 & CustomerSat & 1811 & 10 & 5 & Yes & R-bayesm & \cite{RoAlMc05}\\
\hline \multicolumn{8}{|c|}{Responses to a satisfaction survey for a product}\\ \hline
38 & Avalanches & 394 & 9 & 5 & No & Margherita Maggioni & \cite{Maggioni04}\\
\hline \multicolumn{8}{|c|}{Avalanche frequencies by size and other factors for mapping release areas}\\ \hline
39 & Decathlon & 2580 & 10 & 6 & No & R-GDAdata & (iii) \\
\hline \multicolumn{8}{|c|}{Points per event of decathlon athletes}\\ \hline
40 & Alcohol & 125 & 10 & 5 & Yes & UCI & \cite{ALJY20}\\
\hline \multicolumn{8}{|c|}{Five types of alcohol classified by QCM sensor data}\\ \hline
41 & Augsburg & 95 & 11 & 3 & No & See reference & \cite{TheUrb09}\\
\hline \multicolumn{8}{|c|}{Tax data for districts of the city of Augsburg before and after Thirty Years War}\\ \hline
42 & ImageSeg & 2310 & 16 & 7 & Yes & \verb|www.openml.org| & \cite{UCI17}\\
\hline \multicolumn{8}{|c|}{$3\times 3$ pixel regions of outdoor images classified as object}\\ \hline
\end{tabular}
\label{tdata2}
\end{table} 
A zip file with all data sets in the form in which they were analysed in the present study (i.e., after all pre-processing) is planned to be provided as online supplement 
of the published version of this article. Where a ``true'' clustering is given this is in the first variable.
All variables were scaled to zero mean and unit variance before clustering, except where stated in the following list, which gives information about data pre-processing where this was applied.
\begin{description}
\item[2 Dortmund] The original data set has 203 variables, many of which are not of much substantial interest, with several linear dependencies. The version used here is described in \cite{CorHen16}. 
\item[4 Vowels] The original data set is split into test and training data for
supervised classification. Both are used together here.
\item[5 Bats] The used data set is a preliminary version of what is analysed in \cite{Zamora16} that was provided to me for testing purposes by Veronica Zamora-Gutierrez. A small number of missing values were imputed by mean imputation.
\item[7 OliveOil] The original data set contains classification by 9 regions, which are subclasses of 3 macro areas. The regions were used as ``true'' clustering.
\item[9 Tetragonula] This data set originally contains categorical genetic information. The version used here was generated by running a four-dimensional multidimensional scaling on genetic distances as proposed by \cite{HauHen10}. The resulting data were not scaled before clustering; the original scales represent the original distances.
\item[15 Letters] This data set has originally 20,000 observations, which is too big for handling a full distance matrix. Only the first 2,000 have been used.    
\item[16 Bundestag] The data set was not scaled before clustering. The unscaled version represents comparable voter percentages.
\item[19 StoneFlakes] A small number of missing values were imputed by mean imputation.
\item[21 London] The data were retrieved from the website \verb|maps.met.police.uk/tables.htm| in December 2015. The website has been reorganised in the meantime and the original data are probably no longer available there, however more recent data of the same kind is available. Only the major crime categories were used, divided by the total number of offences; the total number of offences was used as a variable divided by the number of inhabitants. After constructing these features, variables were scaled (the number of serious crimes is 
very low, and not scaling the relative numbers would have strongly reduced their influence on clustering).  
\item[24 MachineKnow] A classification variable is provided, but this was not used as ``true'' clustering here, because according to the documentation this was constructed from the data by a machine learning algorithm, and does not qualify as ``ground truth''.
\item[26 RNAYan] This is originally a data set with $p\gg n$. Unscaled principal components were used as explained in \cite{BatHen20} in line with some literature cited there.
\item[27 RNAKolo] This is originally a data set with $p\gg n$. Unscaled principal components were used as explained in \cite{BatHen20} in line with some literature cited there.
\item[28 Cardiotocography] A variable with less than four distinct values has been removed.
\item[33 SteelPlates] Three variables with less than four distinct values have been removed.
\item[34 BostonHouse] This was originally a regression problem. The original response variable ``housing price'' is used together with the other variables for clustering here. A binary variable has been removed.
\item[35 Ionosphere] Two binary variables have been removed.
\item[38 Avalanches] On top of the first six variables, which give geographical information, the original data has frequencies for avalanches of ten different sizes, categorised by what percentage of the release areas is covered. This information has been reduced to the three variables ``number of avalanches'', ``mean coverage'' and ``variance of coverages''.
\item[39 Decathlon] Only data from the year 2000 onward are used, in order to generate a data set of manageable size. Variables were not scaled, because the decathlon points system is meant to make the original points values comparable.
\item[41 Augsburg] For four count variables the meaning of missing values was ``not existing'', and these were set to zero. Some other missing values were imputed by mean imputation.
\item[42 ImageSeg] Three variables with less than five distinct values have been removed.
\end{description}


\end{document}